\documentclass[10pt]{article}
\usepackage{color,amsmath,amsfonts,amssymb,a4,epsfig}

\newcommand{\R}{{\mathbb{R}}}
\newcommand{\C}{{\mathbb{C}}}
\newcommand{\Z}{{\mathbb{Z}}}
\newcommand{\N}{{\mathbb{N}}}
\newcommand{\E}{{\mathbb{E}}}
\newcommand{\D}{{\mathbb{D}}}

\def\ha{\frac{1}{2}}

\def\pa{\partial}
\def\ra{\rightarrow}
\def\ga{\alpha}
\def\gb{\beta}
\def\gc{\gamma}
\def\gd{\delta}
\def\ge{\varepsilon}

\def\gf{\varphi}
\def\gg{\gamma}

\def\gl{\lambda}
\def\go{\omega}

\def\OPD{${\rm \Psi}$DO}
\newcommand{\opeps}{\operatorname{Op}_{\ge}}
\newtheorem{defi}{Definition}[section]

\newtheorem{lemm}{Lemma}[section]
\newtheorem{prop}{Proposition}[section]
\newtheorem{rem}{Remark}[section]
\newtheorem{coro}{Corollary}[section]
\newtheorem{theo}{Theorem}[section]
\newtheorem{exem}{Example}[section]
\newenvironment{demo}{\noindent {\it Proof.--}
      \begin{quotation}\noindent}{\end{quotation}\hfill$\square $}

\begin{document}

\title{Semiclassical analysis and  passive imaging}
\author{Yves Colin de Verdi\`ere \footnote{Institut Fourier,
 Unit{\'e} mixte
 de recherche CNRS-UJF 5582,
 BP 74, 38402-Saint Martin d'H\`eres Cedex (France);
\textcolor{blue}{\tt http://www-fourier.ujf-grenoble.fr/$\sim $ycolver/}}}

\maketitle

\begin{abstract}
The propagation of elastic waves inside the Earth
provides us with informations about the geological 
structure of the Earth's interior.
Since the beginning of seismology, people have been 
using  waves created by earthquakes or by artificial
explosions. They record the waves as  functions of time using
seismometers located at different stations on the Earth's surface.
Even without any  earthquake or explosion,
a weak signal is still recorded which has no evident structure:
it is a "noise".
How to use these noises? This is the goal of the method of "passive imaging".
The main observation is the following one:
the time correlation of the noisy fields,
computed from the fields recorded 
 at  the points $A$ and $B$, is "close"    to the Green's function
$G(\tau, A,B)$ of the wave propagation. 
 The aim of this paper is to provide
a mathematical context for this approach and to show, in particular,
how the methods of semiclassical analysis can be be used in order to
find the asymptotic behaviour of the correlations.

\end{abstract}

\section*{Introduction}

Seismologists want to recover the physical parameters of the
Earth's interior from  records (called
seismograms),  at the Earth's surface, 
of the elastic waves propagating in its interior. From the mathematical
point of view, it is an example of a so-called "inverse problem'':
the most famous inverse problems are the Calder\'on problem (recovering
the conductance of a domain from  boundary measurements)
and the Kac problem (recovering a Euclidean domain from the spectrum
of its Dirichlet Laplacian).
The use of elastic waves created by an earthquake (resp. an 
artificial explosion)  has been quite successful in order to recover
the large scale structure (resp. the small scale structure, e.g. for oil
detection).
The geological structure of the Earth crust up to depths of 30-50
kilometres
is much more difficult to know. 

On the other hand, only a small part of the seismograms 
was used: this part corresponds to the propagation 
of well identified body and surface waves: typically
 the P-waves, the S-waves and the  Rayleigh
waves.
The last part (called the "coda") of the seismograms were not used.
This was a long standing question asked by K.~Aki, one of the founders
of modern seismology. 
Quite recently, it was observed by M. Campillo and A. Paul \cite{C-P}
that the
time-correlation of coda waves is closely related to the Green's
function.

Later, M. Campillo and his collaborators showed that the
correlations
of ambient noise already yield the Green's function
\cite{DL,SC,SCSR}.
Similar facts had been known before in acoustics
\cite{LW,WL1,WL2,DLC} and helio-seismology \cite{DJ}.
Several theoretical results were derived in \cite{BGP,RSK,SS}.
Applications to clock synchronisation 
\cite{SRTDHK}  and to volcanic eruptions forecasting
\cite{Br}  have been developed.
The  paper \cite{GS} is a recent survey.
So, let us assume that we can recover the Green's function: how does
it help to solve our inverse problem?
Here enters semiclassics (or ray theory): the
semiclassical behaviour of the Green's function
 contains the ray dynamics. Even more,
the full Green's function is not really needed for that, its  phase
is sufficient. So it is natural to study the problem
of passive imaging using the tools of semiclassical analysis.

The goal of my  paper is to present some mathematical
models for this method of passive imaging.
They will be applicable to any kind of wave, but I have in mind
seismic waves. Semiclassical analysis will be used as a way
to get a geometric approximation to the wave propagation
AND in the modelling of the source noise.
I do not at all pretend that the assumptions of my models are 
realistic from the physical point of view. I just hope that they
will  help  seismologists  by providing another
point of view  and other words.
On the other hand, these models put  light onto several
interesting mathematical problems: mode conversion,
inverse spectral problems, mode-ling of random fields \dots

Things work as follows: let us 
assume that we have a medium $X$ (a smooth manifold) 
and a smooth,
deterministic (no randomness in it) linear wave equation in $X$.
We hope  to recover  (part of) the  geometry of $X$ from the wave propagation on it.
We assume  that there is somewhere in $X$ a source of noise ${\bf f}(x,t)$
which is a stationary random field. 
This source generates, by the (linear) wave propagation, a field ${\bf
  u}(x,t)$ (assumed to be scalar in this introduction). This field ${\bf u}$
is recorded at different points $A,B,\cdots$  on long time intervals.
We want to get some information on the 
propagation of waves from $B$ to $A$ in $X$ from the correlation
\[ C_{A,B}(\tau)=\lim  _{T\ra +\infty}\frac{1}{T}\int _0^T {\bf u}(A,t)
 {{\bf \bar{u}}(B,t-\tau)} dt~,
\]
which can be computed numerically from the  fields recorded 
at $A$ and $B $.
In seismology the medium $X$ will be the Earth, 
the waves are elastic waves, the main source noise is due to the
interaction between the ocean or the atmosphere
with the Earth's crust, and the field is recorded at the Earth's surface.

It turns out that $C_{A,B}(\tau) $ is closely related to the
deterministic {\it Green's function} of the wave equation in $X$.
It means that one can hope to recover, using Fourier analysis,
 the propagation speeds of waves
between $A$ and $B$ as a function of the frequency, and, more precisely,
the Hamiltonian or, more generally, the so-called {\it dispersion relation.}

If the wave dynamics is time reversal symmetric,
the correlation  admits also a symmetry by
changing $\tau $ into  $-\tau$; this observation
has been used for  clock synchronisation, see \cite{SRTDHK}.

The goal of this paper  is to give precise formulae for
$C_{A,B}(\tau) $ in the high frequency limit, under some
assumptions on the source ${\bf f}$.
We will assume that the source field is a stationary ergodic 
random field, the covariance of which $K(x,y,s-s')=\E (
f(x,s)\bar{f}(y,s')) $
is given by the Schwartz kernel
of a pseudodifferential operator. This is a strong assumption,
which implies in particular a  
rapid decay of the correlations of the source ${\bf f}$ away from
the diagonal.

More precisely, we  have two  small parameters, one of them entering
into the decorrelation distance of the source noise, the other one 
in the high frequency propagation (the {\it ray} method).
 The fact that both are of the same
order  of magnitude is crucial for the method.
As a result, the method works well in certain frequency ranges of the
wave propagation.
From the physical point of view, the main result
is that, with some assumptions on the support
of the source noise, one can recover the dispersion
relation (the ray dynamics) in some  frequency interval.
A mathematical statement will be that, for $\tau >0$,
  $C_{A,B}(\tau) $ is close to  the Schwartz
kernel of 
$ \Omega (\tau )\circ \Pi  $ where $\Pi $
is a suitable {\it pseudodifferential operator} (\OPD for short),
the principal symbol of which can be explicitly computed,
and $\Omega (\tau )$
is the  (semi)group of the (damped) wave propagation.
This closeness is in general only meant in the $L^2$ sense, but
can be pointwise if the damping is strong  enough.
It implies that we can recover the dispersion relation,
i.e. the classical dynamics,  from the knowledge of
all two-points correlations.

The waves are recorded at the boundary, it implies that
the main part of the recorded noisy fields is coming
from surface waves.
So we end up with another inverse problem: 
how does the dispersion relation of the elastic surface waves
determine the structure of the Earth's interior?
We will discuss the effective (dispersive) ray dynamics
of guided surface waves. The corresponding ideal
inverse spectral problem has been solved in \cite{YCdV5}
from which we recall the main result.

I guess that our results can be interesting because
they provide a quite general situation in which the source
is not a white noise and for which we can still recover
the most interesting part of the Green's function
(the phase). 

Let us also mention on the technical side that, rather than  using  mode 
decomposition, we prefer to work  directly with the dynamics; 
in other words,  we need a {\it time dependent} rather than
 a  {\it stationary}
approach. Mode decomposition is often useful, but it is
of no much help
for a general system with no particular symmetry.

For clarity, we will first discuss the nonphysical case
of a first order wave equation like the Schr\"odinger equation, then the
case of more common wave equations (acoustic, elastic).

In order to make  the paper  readable  by a large audience,
we have tried to make it self-contained by including
sections on pseudodifferential operators and on random fields.
\vskip0.5cm 
{\centerline{ \bf Contents}}
\begin{itemize}
\item 
In Section \ref{sec:general}, we start with a quite general setting
and discuss a general formula for the correlation (Equation
(\ref{equ:cov_op})). This formula will
be made  more precise in several cases in the following Sections.
\item The goal of Section \ref{sec:examples} 
is to introduce  the basic examples which we will discuss in the paper.
\item 
In Section \ref{sec:random}, we introduce a large family of anisotropic random
fields and show the relation between their power spectra and
their  Wigner measures. They are build from white noises
using the Radonification process: our random fields will be the
images of some white noises by some suitable linear operators.
\item 
Section \ref{sec:1comp} is devoted to 
the case of a scalar field driven by a first order in time
differential equation.
This case is quite academic, but will be used later for standard wave equations.
We will discuss first the (nonrealistic) case where the source
is a white noise, then the semiclassical case assuming 
the absence of mode conversion.
\item 
In Section \ref{sec:Ncomp},
we discuss the case of multicomponent
wave equations: we will discuss first the (nonrealistic)
 case where the source
is a white noise, then the semiclassical case using a reduction 
to the case of scalar fields.
\item In Section \ref{ss:coda},
 going back to the beginning of the story, 
we discuss the question of the  correlations
of codas in the framework of ``quantum chaos''.
\item 
In Section \ref{sec:surface}, we focus on the case 
of seismology and discuss
the remarkable fact that the correlation of the
surface waves is enough to image the Earth's crust,
using a waveguide model of the crust.
\item 
Finally, there are 
four  appendices:
\begin{itemize}
\item Appendix A on pseudodifferential operators,
\item Appendix B on classical dynamics, dispersion relations
and generating functions,
\item Appendix C on Egorov Theorem for long times,
\item Appendix D on time averages versus ensemble averages, i.e.
ergodicity of the source field. 
 \end{itemize}
 \end{itemize}

As can been seen in the paper,several interesting
mathematical
problems come out of our discussions; let us mention  
\begin{itemize}
\item Mode conversions associated with an eigenvalue going into the
continuous spectrum (Section \ref{ss:sturm}).
\item Extension of the large time Egorov Theorem, with
Fourier integral operators replacing \OPD's, in order
to be able to properly discuss the case of a source noise
localized on a submanifold.
\item The large time behaviour of the solution of a wave
equation with initially localised Cauchy data, like in an earthquake
(Section \ref{ss:coda}). 

\end{itemize}

Acknowledgement: I would like to thank  Michel Campillo
and his colleagues from LGIT for discussions and collaborations.
I also thank the referee for his great help in order
to improve the initial manuscript.

\section{A general formula for the correlation} \label{sec:general}
{\it \small 
In this quite mathematical Section,
 we present the general context and a general formula
for the correlation which makes quite explicit the relation 
with the Green's function: we will do that on the level of operators
and on the level of  more concrete functions, the integral
(or Schwartz) kernels of the operators. Our model is quite general: it
includes
linear wave equations with some attenuation
and a source noise on the right hand side
of the equation. }

We will   consider the following general damped wave equation:
\begin{equation}\label{equ:PI}
  \frac{d{\bf u}}{dt}+\frac{i}{\ge}  \hat{H} {\bf u}= {\bf f}
\end{equation}
\subsection{Operators  level} \label{sec:operator}
We make the following assumptions:
\begin{enumerate}
\item The parameter $\ge >0$ can be either fixed (say $\ge =1$)
or can be a semiclassical parameter going to $0^+$, like 
$\hbar $ in the Schr\"odinger-like case, or the inverse of the
frequency for the wave equation.
\item The  operator $i\hat{H}/\ge $ is the infinitesimal 
generator of a strongly continuous group (see \cite{Yo}, Chap. IX)
\[ \Omega (t)="{\rm exp}(-it\hat{H}/\ge )" \]
acting on some Hilbert space ${\cal H}$.
\item $\Omega (t)$ is a contraction group: there exists $T_{\rm
    att}>0$,
called the {\it attenuation time}, and $C>0$ so that the 
operator norm of $\Omega (t)$ satisfies
\[\forall t\geq 0,~
\| \Omega (t) \| \leq Ce^{-t/T_{\rm att}}\]
\item The source noise ${\bf f}$ is a stationary (see Section
\ref{ss:random_stat})  random 
field  ${\bf f}:\R \ra {\cal H}$.
We assume that $\E (\| {\bf f}(0) \|^2)<\infty$(\footnote{$\E $
denotes the ensemble average or   expectation value})
and the  covariance of ${\bf f}$ is given by(\footnote{We use
Dirac's bra-ket notation, $ |f> $ is a vector,
$<f| $ is the Hermitian scalar product with $f$,
so that $|f><g| $ is the rank one linear operator:
$x\ra <g|x> f $})
\[ \E \left(~ |{\bf f}(s)><{\bf f}(s')|~ \right) =K(s-s') \] 
with $K(t )\in {\cal L}({\cal H})$(\footnote{${\cal L}(E,F)$ is the
  space of continuous linear maps from the Banach space $E$ into
the Banach space $F$ with the norm topology
 and  ${\cal L}(E):={\cal L}(E,E)$
}), a continuous
 function of $t$ with compact support $\{ |t|
\leq t_0 \}$~(\footnote{The assumption on compact support could be replaced by 
some suitable decay in $t$. It means that the source noise is
uncorrelated for large enough  times}). 
\end{enumerate}

\begin{defi}
The {\rm causal solution} of Equation (\ref{equ:PI}) is
 the solution given
by $u={\bf G}f$, with  ${\bf G}$ a linear operator, satisfying
$u(.,t)=0$ if $f(.,s)$ vanishes for $s\leq t $.
\end{defi}

For physical reasons, it is appropriate to take for
 ${\bf u}$ the causal solution of Equation~(\ref{equ:PI}).
The field  ${\bf u}$ is then stationary, and given by 
\[ {\bf u}(t)=\int _0^{\infty} \Omega (s){\bf f}(t-s) ds ~.\]
 Using Assumptions 3. and   4, one can show that ${\bf u}(t)$  is almost surely defined,
and its   covariance,  defined by
\[ C( \tau )= \E \left( |{\bf u}(0)><{\bf u}(-\tau )| \right) ~,\]
is a continuous linear operator on $\cal{H}$. 

\begin{theo}\label{theo:cov_operator}
The covariance $ C( \tau )$ is given by
\begin{itemize}
\item 
 for $\tau > t_0 $,  
\begin{equation} \label{equ:cov_op}
 C(\tau)=\Omega (\tau)\Pi\quad  
{\rm with}\quad
\Pi= \int _0^\infty \Omega (s) {\cal L}\Omega
^\star (s) ds \quad{\rm and}\quad  
 {\cal L}=\int _{-t_0}^{t_0} \Omega (-u)K(u) du\,,
 \end{equation}
\item for $\tau <0 $, 
\[ C(\tau)=\left( C(-\tau) \right)^\star ~.\]
\end{itemize}
\end{theo} 
Assuming  ${\bf f}$ to be ergodic (see  Section \ref{ss:random_stat}),
the covariance $C(\tau )$ is  also given
almost surely by the {\it time average:}
\[ C(\tau)=\lim _{T\ra \infty }\frac{1}{T}\int _0^T
|{\bf u}(t)> <{\bf u}(t-\tau) | dt ~.\]
In order to perform effective passive imaging, people use
the previous formula for a large enough $T$ and ... computers.

\subsection{Kernels level}
In order to have concrete functions and not only distributions,
we will make some smoothness assumptions: 
\begin{itemize}
\item $X$ is a {\it smooth compact manifold} of dimension $d$
with a smooth measure $|dx|$. For applications to seismology
$X$ will be the Earth or (in Section \ref{sec:surface}) the Earth's surface. 
\item ${\bf u}(x,t),~x\in X,~t\in\R $ is  {\it the field }(scalar or vector
valued)
with values in $\C^N$ (or $\R^N$).
\item The generator of the group
$\Omega (t)$ is an elliptic (see Definition
\ref{def:admi}) differential (or
 pseudodifferential) linear operator 
$\hat{H}$  acting with a dense domain  of a Hilbert space
 ${\cal H}\subset {\cal D}'(X,\C ^N)$.
\item The  {\it source} $ {\bf f}$ is a smooth 
 {\it stationary  random field }
on $X\times \R $ with values in $\C^N$ (or $\R^N$).
Its matrix valued correlation kernel
is the smooth function on $X\times X\times \R$, with values
in ${\rm End}(\C ^N)$, given by 
\begin{equation} \label{equ:correl_source}
 \E ( {\bf f}(x,s)\otimes  {\bf f}^\star (y,s'))
=K(s-s',x,y) ~.\end{equation}
\end{itemize}
\begin{defi}\label{defi:prop}
The {\rm propagator} or {\rm Green's function}
 $P$ is the Schwartz kernel of $\Omega(t)$
defined 
by: 
\[ (\Omega(t){\bf v})(x)=\int _X P(t,x,y){\bf v}(y) |dy|~. \]
\end{defi}
The propagator $P$ satisfies
\[ \int _X P(t,x,y)P(s,y,z)|dy| =P(t+s,x,z) \]
which comes from the group relation
$\Omega(t+s)=\Omega(t)\circ \Omega(s)$.
The causal solution of Equation~(\ref{equ:PI}) is then given by
\[ {\bf u}(x,t)=\int_0^{\infty}ds 
 \int_X P(s,x,y){\bf f}(t-s,y)|dy| ~.\]
Due to the ellipticity of $\hat{H}$, this is a smooth function. 

The kernel of $C(\tau)$ is defined by:
\begin{equation} \label{equ:correl0}
  C_{A,B}(\tau)=\E ({\bf u}(A,0)\otimes
 {\bf u}(B,-\tau)^\star )  ~,\end{equation}
where $A,B \in X$ and $ C_{A,B}(\tau)\in {\rm End}(\C^N)$.

From Equation~(\ref{equ:cov_op}), we get an explicit
integral formula for $ C_{A,B}(\tau)$
in terms of $P(t,x,y)$ and $K(t,x,y)$;
Equation~(\ref{equ:cov_op})
can be written  for $\tau > t_0$ as follows(\footnote{If $R$ is an operator 
we will denote by $[R](x,y)$ its Schwartz kernel}):
\begin{equation} \label{equ:corr_oper}
 {C_{A,B}(\tau )}= [  \Omega(\tau)\Pi ](A,B) 
\end{equation}

\section{Examples} \label{sec:examples}

{\small \it In this Section, we will present the examples of ``damped wave
equations'' which will be used in this paper.
We start with concrete examples and go then to more  general 
semiclassical examples in the scalar case as well as 
in the multi-component case.}

\subsection{Damped Schr\"odinger equation}\label{ss:schro}

Let $X$ be a smooth compact boundary-less
 Riemannian manifold whose  Laplace-Beltrami  operator is denoted by
$\Delta $.
Let us give $a:X\ra \R $ a smooth non negative  function,
$V$ a smooth
 real valued function on $X$, $\hbar $ a non negative
{\it constant},  and
$ \hat{K}=-\hbar^2 \Delta + V(x) $, and take:
\[ \frac{\hbar}{i}(u_t +a(x)u)  + \hat{K}u= g~. \]
It is a particular case of Equation (\ref{equ:PI})
where $\ge=\hbar$,  $\hat{H}= \hat{K}+ \frac{\hbar}{i} a(x)  $,
 $f=\frac{i}{\hbar}g$ and the Hilbert space is $L^2(X)$.

 Let us note for future use that, if $\hbar \ra 0$, 
the principal symbol (see Appendix A, Section \ref{sec:basic})
 of our equation reads
$\omega + \| \xi \|^2 + V(x)$ where our (extended) phase space
is $T^\star (X\times \R )$ with canonical coordinates
$(x,\xi,t,\go)$
and $a(x)$ is a subprincipal term entering in the
transport equation, but not in the classical dynamics (see Appendix A).
The attenuation time can be taken as
$T_{\rm att}=1/ \inf_{x\in X} a(x)$.
Let us note that this equation has no physical meaning, it
is just a kind of toy model.

\subsection{Damped wave equations}\label{ss:wave}
As in the previous Section, $X$ is a compact Riemannian manifold.
Let us start with 
\begin{equation}\label{equ:wave}
 u_{tt}+a u_t -\Delta u= f,   \end{equation}
($a$ is a  smooth function which satisfies
$\inf _{x\in X} a(x)>0 $). It
corresponds to Equation~(\ref{equ:PI}) if we set
\[ {\bf u}=
\left( \begin{array} {c}\ge \sqrt{-\Delta} u \\ -i\ge u_t \end{array} \right),~
{\bf f}:=\left( \begin{array} {c}0  \\-\ge^2 f \end{array} \right)~
 \]
and
\[ \hat{H}=
\left( \begin{array} {cc} 0 &-\ge \sqrt{-\Delta} \\
 -\ge\sqrt{-\Delta} & -i\ge a
 \end{array} \right)
 ~.\]
The Hilbert space is defined by the energy norm
$\| (u_1,u_2) \|^2 = \int_X \left(|u_1|^2+ |u_2|^2
\right)|dx| $.

\begin{rem}It follows from \cite{R-T} that,
if  $a(x)\geq 0 $ everywhere and there exists some $T_0 >0$
so that  the support of $a(x)$ crosses every geodesic arc of length 
$\geq T_0$, then $\Omega (t)$ is 
a contraction group.
\end{rem}

One can replace the very simple wave equation (\ref{equ:wave})
by more sophisticated multicomponent wave equations
like the equation of elastic waves.
The general form will be:
\begin{equation} \label{equ:wave_gen}
u_{tt}+ A u_t -\Lambda u =0~,
\end{equation}
where $u(x,t)\in \C^{N}$, $\Lambda $ is an elliptic system
of degree $2$ and $A$ is a differential  operator
of degree $0$  representing the
attenuation.

\subsection{Pseudodifferential equations}\label{ss:scalar}

{\it \small This example will be the subject of Section \ref{sec:1comp}.}

This is a generalisation of the damped Schr\"odinger equation
(Section \ref{ss:schro});
we now introduce a small parameter $\ge $
which, in the case of the damped Schr\"odinger equation,
is Planck's constant $\hbar $.
The  semiclassical regime will correspond to $\ge \ra 0$,

We  assume that the dynamics is generated by a
pseudodifferential
operator  (a \OPD, 
 see Appendix A).
Our equation looks then like:
\[ {\bf u}_t +\frac{i}{\ge}\hat{H}_\ge {\bf u}= {\bf f} \]
with
\[ \hat{H}_\ge ={\rm Op}_\ge (H_0+\ge H_1) ~.\]
Here $H_0$ is real valued and elliptic,
 and is the Hamiltonian of the ray dynamics,
while $\Im H_1 <0 $ provides the attenuation. 

In the case of the damped Schr\"odinger equation (Section
\ref{ss:schro}), we have $\ge=\hbar$, 
$H_0= \| \xi \|^2 +V(x)$ and $ H_1=-ia(x) $.

\subsection{Pseudodifferential systems}\label{ss:systems}
{\it \small This example will be the subject of Section
  \ref{sec:Ncomp}.}

The system is of the following form
\[{\bf u}_t +\frac{i}{\ge}\hat{H}_\ge {\bf u}= {\bf f}~,\qquad\text{with}\qquad 
\hat{H}_\ge ={\rm Op}_\ge (H_0+\ge H_1) ~.\]
Here
\begin{itemize}
\item ${\bf u}(x,t)\in \C^N$
\item
 $H_0(x,\xi)$ is an Hermitian  symmetric  matrix.
Each  eigenvalue $\gl(x,\xi) $ 
of $H_0(x,\xi)$  gives a ray dynamics associated to a
polarised wave. The polarisation bundle is the associated
eigenbundle $E_\gl (x,\xi)$
\item 
 $H_1(x,\xi)$ is a complex  matrix. 
 Let us introduce the quadratic forms
$q_\gl (x,\xi)$
which is the restriction of $\langle \Im H_1 x|x \rangle $
to $E_\gl(x,\xi)$.
The condition $q_\gl <0$ 
ensures the attenuation. 
\end{itemize}
 This  includes
\begin{itemize}
\item 
The high frequency limit of the wave equations (\ref{equ:wave}):
we have
\[ H_0=\left( \begin{array}{cc} 0& -\| \xi \| \\
-\| \xi \| & 0  \end{array} \right), ~
H_1=\left( \begin{array}{cc} 0& 0 \\
0 & -ia   \end{array} \right)~. \]
A similar transform can be used for more general wave equations,
like the elastic wave equation with $N=6$.
\item An effective pseudodifferential
wave equation associated with stratified media.
 They are usually \OPD's with a nontrivial 
dispersion relation (see Section \ref{sec:surface}).  
\item Phase space  dependent damping included in $H_1$:
 this is usually the case for
seismic waves. The attenuation can be created by the small-scale
physics which is not uniform in general. 
\end{itemize}

\section{Random fields and semiclassics }
\label{sec:random}
{\small \it 
Our aim in this section is to build quite general random fields 
with  correlation distances given by a small parameter $\ge$.
It seems to be  natural for that purpose to use $\ge-$pseudodifferential
operators. We will see how to compute the generalised power 
spectrum using Wigner measures. We will not restrict ourselves to
time-dependent
random fields. The  subsection \ref{ss:random_stat}
is devoted to the time-dependent case
and we introduce there the notions of stationarity and ergodicity.}
\subsection{Covariance of Random fields}
A random field is simply a random variable with values in some
functional space, like a Hilbert space ${\cal H}$.
We will consider
random fields $f=f(\bar{\go} )\in {\cal H}$ where $\bar{\go}$
belongs to $ \Omega$,
the  sample probability space.
We will assume that $f\in L^2(\Omega , {\cal H})$ and, 
  without loss of generality,
that our fields have mean $0$:
$\E (f)=0 $.

The {\it covariance} of $f$ is the operator
$C=\E ( ~|f><f|~)$ on ${\cal H}$.
This operator is {\it Hermitian and positive.}
${\cal H}$ will often be a space of functions, and
we will consider the Schwartz kernel $[C](x,y)$ of $C$.

If ${\cal H}$ is finite dimensional, it admits
random fields for which the covariance is the 
identity operator. We call them {\it white noises}.

\subsection{Cylindrical  random variables}
 {\small \it For this section, see \cite[Part II]{Schwartz} and \cite[Chap. IV]{GV}}
\begin{defi}
A {\rm cylindrical random variable} $f$ on the Hilbert space  ${\cal H}$ 
is given by the following data:
for each continuous linear map $A: {\cal H} \ra E $
with $\dim E <\infty$, let $f_{A} $ be a random variable with values in  $E$.
These random variables must satisfy the following compatibility
condition: if  $l:E \ra F $ is linear, $f_{l\circ A}=l\circ  f_{A} $ (at the informal
level, we can write $f_A=A\circ f$).
\end{defi}

In particular, if $g:{\cal H} \ra \C $ is of the
form $g=F\circ A $ with $A:  {\cal H} \ra E$ linear, 
$\dim E <\infty$ and $F\in L^\infty (E)$, we can
define $\E (g):=\E (F\circ f_{A})$.

\begin{exem} Let $ {\cal H}$ be  $L^2(\R/2\pi \Z )$,
$\Omega = \{ -1, +1 \}^\Z $, with the product measure
$\Pi \left( \ha (\delta (-1)+\delta (1) )\right)$.
Then  $f_\omega(x)=\sum _{n\in \Z}\omega_n\, e^{in x}$ is a cylindrical random
variable on ${\cal H}$.  
$f$ is not a random variable with values in $ {\cal H}$,
but it is a random Schwartz distribution
in the Sobolev space $H^s (\R/2\pi \Z )$ for $s<-\ha $. Indeed, for any $\omega\in\Omega$ we have
$\|f_\omega\|^2_{H^s}=\sum_{n}(1+|n|^2)^s<\infty$.
\end{exem}

\begin{defi} The {\rm Fourier transform}
$\widehat{w} $  of a  cylindrical random variable  $w$
is a function on ${\cal H} $ defined as follows.  For each $x\neq 0$, one can
take $\Pi_x$ the orthogonal projector on the one-dimensional subspace $\mathbb{C}x$. To this projection is associated
the random variable $w_{\Pi_x}$ taking values in $\mathbb{C}x$. The value of the Fourier transform
at $x$ is given by $\hat{w}(x)=\E({\rm  exp}(-i\langle w_{\Pi_x} |x \rangle ))$. Informally, we can write 
$\hat{w}(x)=\E ({\rm  exp}(-i\langle w |x \rangle ))$.
\end{defi}

To any random variable $f$
 with values in  ${\cal H} $
is associated the cylindrical random variable  defined
by $f_A=A\circ f$.
If the cylindrical random variable  comes from a random variable $f$
 with values in  ${\cal H}$,
 this  random variable is unique and we will say that
our cylindrical random variable ``is'' a random variable.

We have the important
\begin{theo}[Sazonov   \cite{Schwartz}]\label{theo:sazo}
If the Fourier transform $\hat{w}$
of the cylindrical  random variable  $w$  is continuous 
and $A: {\cal H}\ra  {\cal K}$ is  {\rm Hilbert-Schmidt\footnote{A
 Hilbert-Schmidt operator $A$ is an operator
whose Schwartz kernel $[A](x,y)$ is in $L^2(X\times X)$ and the
Hilbert-Schmidt norm  $\| A  \| _{\rm H-S}$ of $A$ 
is the $L^2$ norm of $[A]$.}     ,}
then $A(w)$ is a  random variable on $ {\cal K}$.
The covariance of $A(w)$ is the operator
$AA^\star $.
\end{theo}

\subsection{White noises in Hilbert spaces}

A white noise $w$ on the Hilbert space $ {\cal H}$
is a cylindrical random variable on  $ {\cal H}$
of mean $0$ and covariance given by the identity:
$$
\forall e,f \in {\cal H},~\E ( \langle e|w \rangle \langle w|f \rangle)=
\langle e|f \rangle\,.
$$

\begin{defi}
The {\rm Gaussian } white noise is the white noise on  $ {\cal H}$
whose Fourier transform is  ${\rm exp}(-\| x \|^2)$.
\end{defi}

Sazonov's Theorem \ref{theo:sazo} applies in particular to
the {\it Gaussian white noise}.

If the Hilbert space ${\cal H}$ is not finite dimensional,
the white noises are
 ``random distributions'' in the following sense:
 if $ {\cal H}=L^2(X)$,
we can take for ${\cal H}$ a Sobolev space $H^s$
 of negative order $s<-\dim X /2$ and for $A$ the injection
from $L^2(X)$ into $H^s(X)$.
So that $w$ can be viewed as a random distribution in $H^s(X)$.

\subsection{An important example} 

Take a self-adjoint operator $\hat{H}$  on ${\cal H}$ with compact
resolvent. Let us denote by $A_I$ the  (finite rank) spectral projection 
onto a compact interval $I$.
The random field $f_I=A_I w$ can also be defined
in terms of the spectral decomposition $(f_j,\gl_j)$ of $\hat{H}$.
$f_I=\sum_{\gl_j\in I} a_j(\bar{\go}) f_j $
where the coefficients $a_j$ satisfies
$\E (a_i)=0$ and $\E (a_i \overline{a_j})=\gd _{i,j}$.

\subsection{Stationary Random fields}\label{ss:random_stat}

We  will now consider {\it time dependent} random fields
$f: t\ra f(t) \in  {\cal K}$ with  $t\in \R$.
Equivalently, our global Hilbert space is now
${\cal H}=L^2(\R, {\cal K})$. 
We can easily define an action of $\R$ on such a random field.
For $T \in \R$, we define $T.f$ by  $(T.f)(t)=f(t-T)$.
We will say that 
\begin{defi}
$f $ is {\rm  stationary} if there exists  a group of  
probability preserving transformations on $\Omega $, denoted by $\{X_T,\ T\in\R\}$,
so that 
$T.(f(\bar{\go}))=f(X_T(\bar{\go}))$.
\end{defi}

Let us consider the time dependent  fields
$f:\R \ra  {\cal K}$ as elements of
a topological vector  space ${\cal B}$.
Stationarity  corresponds to the physical intuition
that $T.f$ is another random choice of the  function $f$
(see also  Wiener's book \cite{Wi}).

If $f$ is a stationary random field,
the covariance kernel
$\tilde{K}(t,t')=\E (~|f(t) > <f({t'})| ~)=K(t-t')$, where $K(s)\in {\rm Herm}({\cal K})$.

\begin{defi}
A  stationary random field $f$ is {\rm ergodic}
if any measurable subset of $\Omega $, invariant by 
the transformations $X_T$, is of probability $0$ or $1$.
\end{defi}

We have then the celebrated
\begin{theo}[Birkhoff]
If $f$ is  stationary and ergodic
and $F:\Omega  \ra \R $ is integrable with respect to the probability 
measure, then, for almost every $\bar{\go}\in \Omega $, 
\[ \lim _{T\ra \infty }\frac{1}{T} \int_0^T F(X_t (\bar{\go})) dt =
\E (F)~.  \]
\end{theo}

\subsection{Examples of stationary fields}

We will start with a Gaussian white noise $w$ on
$L^2(\R, {\cal K})$.
Let us consider a Hilbert-Schmidt operator $A:L^2(\R, {\cal K})
\ra  {\cal K}$. Let us introduce the stationary random
 field $f$ defined by
\[ f(t)=A(t.w) ~.\]
If the Schwartz kernel of $A$ is
$a(s,x,y)$ we have, formally, 
$$f(t,x)=\int _X  |dy| \int _\R ds a(s+t,x,y) w(s,y)\,.
$$

Le us assume that 
$Aw =\int _\R \tilde{A}(s)w(s) ds $
with $\tilde{A}(s) $ a Hilbert-Schmidt
operator smoothly depending on $s$, with compact support $\subset [-t_0/2, t_0/2]$.
From the formula in Theorem \ref{theo:sazo},
 the correlation operator
is 
\begin{equation}\label{equ:cor_stat}
  K(t)=\int _\R \tilde{A}(t+s)\tilde{A}(s)ds ~\end{equation} 
which is supported in  $[-t_0, t_0]$.

\begin{exem} {\rm Stationary scalar noise on the real line:}
let us take ${\cal K}=\R$,
 $F\in C_o^\infty (\R,\R)$\footnote{If $X$ is a smooth manifold
$C_o^\infty (X,\R)$
denotes the space of smooth functions with compact support} and 
the operator $A$ given by 
$w\ra \int _\R  F(-s)  { w}(s) ds  $.
Then 
$f$ is the convolution $F\star w$.
If we take the Fourier transform w.r.to time, we find for each frequency $\go\in\R$
$\hat{f}(\go)=\hat{F}(\go)\hat{{ w}}(\go) $, so that 
$\E (| \hat{f}(\go)|^2)= |\hat{F}|^2(\go)$.
The positive function $\go\in\R\mapsto|\hat{F}|^2(\go)$ is usually called 
the {\rm power spectrum} of the stationary noise $f$.
\end{exem}

\begin{exem} {\rm Brownian motions:}
if ${\cal K}=\R$ and  ${w}$ is the Gaussian white noise,
the Brownian motion is defined by   $b(t)=\int _0^t { w}(s)ds $,
which 
is in $L^2 ([0,T])$ for all finite $T$.
Notice that $b$ is NOT a stationary random field! 
\end{exem}

\subsection{Random fields from   pseudodifferential operators }
{\it \small The main goal of the present section  is  to build natural 
random fields 
which are  non homogeneous  with small distances
of correlation of the order of $\ge \ra 0$.
The noise is non homogeneous in $X$, but could also
be non  isotropic w.r. to directions.}

We will use as Hilbert-Schmidt operators some
$\ge$-pseudodifferential operators.

\subsubsection{Random fields  from pseudodifferential operators}

 Let us  take for {\it $\ge-$dependent }
random fields  on a manifold $Z$ (which can be $X\times \R$)
the image of a white noise on $L^2(Z)$  by an $\ge$-pseudodifferential operator
${A}$ of smooth compactly supported 
symbol $a(z,  \zeta)\in S^{-\infty}_{\rm class} $
 (see Appendix A, Definition \ref{defi:class-symb});
this operator $A$ is Hilbert-Schmidt.

The correlation $C(z,z')$ will then be given 
as the Schwartz kernel of $A A^\star $
which is a  $\Psi DO$ of principal symbol $|a|^2 $.
We have
\[ C(z,z') = (2 \pi \ge) ^{-d}k\left(z, \frac{z'-z}{\ge }\right)+
O(\ge ^{-(d-1)}),\]
where $k(z,.)$ is the Fourier transform
with respect to $\zeta$ 
of  $|a|^2 (z, \zeta  ) $.

This construction gives smooth random fields which can be localised 
in some very small domains of the manifold $Z$,
which are non isotropic (the covariance kernel $C(z,z')$ for $z'$
close to $z$,  depends
on the direction and  not only on the distance $|z-z'|$) and
which have small distance of correlations of the order of $\ge$.
Moreover it will allow to use techniques of 
micro-local analysis with the small parameter given by
$\ge $. 

\subsubsection{Power spectrum and Wigner measures}
The Wigner measures represent the phase space energy densities of a family
of functions depending on a small parameter $\ge$. 
\begin{defi}
If $f=(f_\ge) $ is a family of functions on $Z$, bounded in $L^2_{\rm loc}(Z)$,
 the {\rm  Wigner measures} $W_f^\epsilon$ of $f$
  are the   measures on the
phase space $T^\star Z$ defined, for $\gf \in C_o^\infty (T^\star Z)$,
by 
(see Appendix A, Definition \ref{defi:opd})
\[ \int \gf  dW_f^\epsilon := \langle  \opeps (\gf) 
f_\ge|f_\ge  \rangle ~. \]
The measures $dW_f^\ge $ are the phase space {\rm densities  of energy}
of the functions $(f_\ge) $.
\end{defi}

We  now define: 
\begin{defi}  The {\rm power spectrum} of the  random field 
$f=(f_\ge)$ is the phase space measure   ${\cal P}_f^\ge $ defined by:
\[ {\cal P}_f^\ge = \E (  W_f^\ge ) ~:\]
the power spectrum of a random field
 is  the   average of its Wigner measures.
\end{defi}

\begin{prop}
 The power spectrum ${\cal P}$
of  $f_\ge= {\rm Op} (a) {w}$,
satisfies:
\[ {\cal P}_f^\ge = (2\pi \ge)^{-d}|a|^2(x,\xi)|dx d\xi | +O(\ge^{-(d-1)}) ~.\]
\end{prop}
\begin{demo} Let us put $A=\opeps  (a)$, we have
\[ \langle  \opeps (\gf )A{w}|A{w} \rangle=
 \langle  A^\star \opeps (\gf )A{w}|{w} \rangle \]
and $\forall B, ~\E (  \langle  B{w}|{w} \rangle )=
{\rm trace}(B)$. We get
\[ \E \left(  \int \gf  dW _f ^\ge \right)=
{\rm trace}(A^\star \opeps (\gf ) A) \]
which can be evaluated using the $\Psi$DO calculus
as
\[  \E \left(  \int \gf  dW _f ^\ge \right) =
 (2\pi \ge)^{-d}\int \gf|a|^2 dxd\xi
+ O(\ge^{-(d-1)}) ~.\]

\end{demo}

\subsubsection{Time dependent random fields}\label{ss:tdrf}
If  $Z=X\times \R $ is the space-time, 
we will denote by $(x,\xi)$ (resp. $(t,\go)$) the canonical local
coordinates  in $T^\star X$ (resp.$T^\star \R$). We will
take our random fields  as before: $f=L_\ge {w}$;
the symbol $l$ of $L_\ge =\opeps (l) $  is assumed to be
 $l( x,\xi,\go )\in S^{-\infty}_{\rm class} ( X \times \R)$
(independent of $t$).
The field $f$ is stationary.

In this case, the correlation is given by:
\begin{equation} \label{equ:psido}
 K(x,y; t)=[ L L^\star](x,y;0,t)\end{equation}
which is the  Schwartz kernel of a  $\Psi DO$ of principal  symbol
$|l|^2   (x,  \xi;\go )$.
\begin{lemm} \label{lemm:cov_stat}
The covariance  operator
$K(t)$ is a \OPD~ in the space $\Psi^{-\infty }_{\rm class}$
of principal symbol
\begin{equation} \label{equ:ppalcor}
\frac{1}{2\pi \ge }\int_\R |l|^2 (x,\xi;\go)e^{it\go /\ge }d\go ~.
\end{equation}
\end{lemm}

\section{Correlation for   1-component wave equation}\label{sec:1comp}
{\small \it In this section, we will study the case of a scalar field
given as in Section \ref{ss:scalar}. First we describe,
as a warming up,  a case where
the correlation is exactly computable, then we compute the asymptotic
of the correlation in the "semiclassical case".}

\subsection{An example of an exact formula for the correlation}
Let us consider the 1-component "wave equation" (with $\ge=1$)
\[ \frac{du}{dt}+ ku + i\hat{H}_0 u =f \]
where $k>0 $ is a  constant and $\hat{H}_0$
is self-adjoint on ${\cal K}$  with a compact resolvent.
In particular 
$\Omega (t)= e^{-kt}U(t)$ where $U(t)$ is the unitary group
$U(t)={\rm exp}(-it\hat{H}_0)$. 
Let us assume that ${\bf f}= \Phi \star A_I w $, i.e.
$f(t)=A_I \left( \int \Phi(t-s)w(s) ds \right) $, 
with $w$ a Gaussian white noise on $L^2(\R,{\cal K})$,
 $\Phi \in C_o^\infty (]-t_0/2, t_0/2[;\R)$ and $A_I=\chi _I (\hat{H_0})$ is 
the  finite rank spectral projection of $\hat{H}_0 $ associated
to the compact interval $I$.
The operator $A: w \ra  \int _\R  \Phi(-s) A_I w(s)ds $
is clearly Hilbert-Schmidt. 
From Equation (\ref{equ:cor_stat}), 
 the correlation is given by 
$K(t)=\Psi (t) A_I$ with 
$\Psi(t)=\int _\R \Phi(t+s)\Phi (s) ds $.
Using Theorem \ref{theo:cov_operator},
we get
\begin{theo}
We  have  the following simple formula for $\tau >t_0$:
\[  {C(\tau )}=\frac{1}{2k}F(\hat{H_0})\Omega (\tau)~,\]
with $F(\go)=\chi _I(\go) {\cal F}_{t\ra \go}(\Psi(t) e^{kt})$(\footnote{$ {\cal F}$
is the Fourier transform
$ {\cal F}(g)(u)=\int {\rm exp}(-ius) g(s)ds $.
}), 
which is an exact relation between the correlation and the propagator $P_0$
of  the wave equation without attenuation.
\end{theo}
\begin{rem}\label{rem:green}  At least formally, we can take
$\Phi=\gd (0), ~I=\R$, then $f=w$ and $F\equiv 1$.
The correlation is then equal to the Green's function up to
the factor $1/2k$.
\end{rem}
\begin{rem} Assuming that $\hat{H}$ is ``real valued''(i.e. 
$ \overline{ \hat{H}f}=\hat{H}\overline{f} $),we have
$C_{A,B}(-\tau)= \overline{C_{A,B}(\tau)}$. 
\end{rem}

\subsection{Semiclassical assumptions}\label{subsec:assumptions}
Let us start with the semiclassical Schr\"odinger like  equation
\[ {u}_t + \frac{i}{\ge} \hat{H}{u}=
 { f} \]
where
\begin{itemize}
\item $ \hat{H}$ is {\bf admissible} (Definition \ref{def:admi}
in Appendix C):
$ \hat{H}$ is an { $\ge-$pseudodifferential operator:}
\[   \hat{H} :=\opeps
(H_0+\ge H_1) \]
with  
\begin{itemize}
\item The {\bf principal symbol} $H_0(x,\xi):T^\star X\ra  \R $,
which gives the classical (``rays'') dynamics, belongs to
$\Sigma _m$, with $m >0 $,   and is  elliptic of degree
$m  $ (see Definition \ref{def:admi}).
\item The {\bf sub-principal symbol} $H_1(x,\xi)$ belongs to
$\Sigma _{m-1}$  and admits some
positivity property which controls the attenuation:
there exists $k>0$, such that
\[h_1(x,\xi):= \Im H_1(x,\xi) \leq -k,\quad \forall (x,\xi)\in T^*X ~.\] 
\end{itemize}
\item The {\bf random field}
 $f$ is given by ${ f} =\opeps (l(x,\xi,\go))w$
with $w$ a white noise on $X\times \R$ and 
with $l$ smooth, compactly supported
w.r. to $(x,\xi) $ and whose Fourier transform w.r. to
$\go $ is compactly supported in $]-T_0,T_0[$. The power spectrum of ${ f}$
is $(2\pi \ge)^{-(d+1)}|l|^2  (x,\xi,\go)$.
The covariance operator $K(t)$
is a \OPD~ given by Lemma  \ref{lemm:cov_stat} and supported in 
the interval $]-T_0\ge ,T_0 \ge [$.
\item We introduce some compact energy interval $I$ 
so that ${\rm Support}(l)\subset H_0^{-1}(I)$.
\end{itemize}

\begin{lemm} \label{lemm:subp}
Under the previous assumption on $H_1$, we can take
for  $T_{\rm att}$ (defined in Section \ref{sec:operator})
 any number $> 1/k$ provided
that  $\ge $ is small enough.
\end{lemm}

\begin{demo}
\[ \frac{ d}{dt} \langle v(t)|v(t) \rangle =
2 \Re \langle v(t)| -i\hat{H}_1(t)v(t) \rangle \]
and we use G{\aa}rding inequality (see \cite{Di-Sj}): if $a\geq 0$,
$\opeps (a) \geq -C $
for any $C>0$ and $\ge $ small enough.
\end{demo}

\subsection{Ehrenfest time}\label{defi:ehren}

\begin{defi}[Ehrenfest time]
Let  $X$ be a smooth  compact manifold of dimension $d$.
Let us consider a smooth elliptic  Hamiltonian $H_0$
on $T^\star X$. We fix also some compact energy interval $I$.
\begin{itemize}
\item The {\rm Liapounov exponent }
$\Lambda _I=\Lambda   \in ]0,+\infty [$ is  {\bf any}  real
number(\footnote{For technical reasons, not taking the infimum
of these $\Lambda$'s is more convenient}) 
 for which the differential of the  Hamiltonian flow $\Phi_t$ of $H_0$
satisfies the following {\rm uniform} estimate:
\[\exists C>0,~ \forall z\in T^\star X {~\rm with~}H_0(z)\in I ,~\forall
t\geq 0,~
\|d\Phi_t (z)\|\leq Ce^{\Lambda  t} ~.\]
\item Let us denote by $\ge  >0 $ the {\it semiclassical parameter}.
The {\rm Ehrenfest time},
$T_{\rm Ehrenfest}\in ]0,+\infty ]$ is defined as
\[T_{\rm Ehrenfest}:=|\log \ge |/\Lambda ~.\]
\end{itemize}
\end{defi}
The intuition behind the definition of the Ehrenfest time is the
following one:
if $C$ is a phase space cell of diameter $\sim \ge$,  
the diameter
of the cell $\phi_t (C)$  remains $\ll 1$ for any time $t$ smaller than $T_{\rm Ehrenfest}$.
We will need a smaller time:
\begin{defi}\label{defi:tgamma} Let us give $0<\gg < \ha  $.
The time $T_\gg=T_\gg (I)$  is defined
as 
\[ T_\gg:=\left(\ha -\gg \right) T_{\rm Ehrenfest}~.\]
\end{defi}

The previous definition will be used via the
\begin{lemm} \label{lemm:diff}
 Assume that we have a flow $\Phi_t$ on a compact manifold $X$
with a Liapounov exponent $\Lambda $, then
we have the following uniform estimate
\[ \forall \ga  \in \N^d,~\forall \Lambda' >
\Lambda,  \qquad |D^\ga \Phi_t (z)| =O (e^{\Lambda '|t||\ga |})~.\]
\end{lemm}

\subsection{Main result ($N=1$)}\label{ss:main-1}
We
get the main result: 
\begin{theo}\label{theo:N=1}
With the assumptions and notations  of Section \ref{subsec:assumptions} 
and   $\gg  <\ha$,
  the correlation
  is given, for $\tau >0$,  by 
$ {C(\tau)}=\Omega (\tau)\circ \opeps  (\pi)+R  $ 
where 

\begin{itemize}
\item
\[ \pi (x,\xi)=\int _0^{T_\gg (I)}  
{\rm exp}\left(2 \int _{-t}^0 \Im (H_1)
 (\Phi_s (x,\xi))ds \right)|l|^2 
\left(\Phi_{-t} (x,\xi),-H_0(x,\xi)\right) dt ~\]
belongs to $S ^{-\infty}_{\ha -\gg} $,
 and is supported in
$H_0^{-1}(I)$.

\item
\begin{enumerate}
\item {\bf [No assumption on the attenuation $k$]} we take
 \[ \gg=\gg_1=\frac{k}{ 2 (\Lambda  +k )}~,\]
and,
for   any  $\ga < 2\gg_1 $, we have the {\rm Hilbert-Schmidt} estimate
\[ \| {R} \|_{\rm H-S}=O(\ge^{\ga - d/2 })~\]
 (the  operator ${\Omega }(\tau )\circ \opeps (\pi)$ is
 Hilbert-Schmidt            
with an
Hilbert-Schmidt norm of the order of $\ge ^{-d/2}$
if $\pi $ is not $\equiv 0$);
\item {\bf [Large attenuation]}  if $k> 2d\Lambda $, we take
\[ \gg=\gg_2= \frac{k-2d \Lambda}{ 2(\Lambda +k)}~,\]
and,
for any $\gb<2 \gg_2 $, 
for $(A,B,\tau )$ generic (see Definition \ref{defi:generic}),  
we have the {\rm point-wise} uniform estimate
 \[  [R](A,B) =
O(\ge ^{\gb -\frac{d}{2}})~.\]
\end{enumerate}
\end{itemize}
\end{theo}

\begin{demo}
From Lemma \ref{lemm:calL}, we know that the operator
 ${\cal L}$ introduced in Theorem \ref{theo:cov_operator}
belongs to $ \Psi^{-\infty}_{\rm class} $
and his   principal symbol is  $|l|^2(x,\xi,-H_0(x,\xi ))$.
The correlation operator
$C(\tau)$ is given (Theorem  \ref{theo:cov_operator}) by 
$C( \tau)=\Omega( \tau)\circ \int_0^\infty \Omega (u){\cal L}
\Omega ^\star(u) du $.
We  split the integral
$\int_0^\infty =\int_0^{T_\gg}+ \int_{T_\gg}^\infty $.
\begin{itemize}
\item The first integral is estimated using the long time 
Egorov Theorem.
\item The second one is  estimated using the exponential
decay of $\Omega (t)$
\end{itemize}

The main term comes from Theorem \ref{theo:long} with the formula for 
$a_0 $ given in  Theorem \ref{theo:egorov}.
We split the remainder $R$ into
$R=R_1+R_2 $
with
\begin{itemize}
\item 
\[ R_1 =\Omega (\tau) \int_0^{T_\gg }r_1 (s) ds \] 
and $r_1(s) \in \ge ^{2\gg  }\Psi^{-\infty }_{\ha -\gg }$
(using Theorem \ref{theo:long})
\item 
\[ R_2 =\int _{T_\gg }^\infty \Omega (s+\tau){\cal L }\Omega ^\star(s)ds 
\]
with ${\cal L }\in  \Psi ^{-\infty }_{\rm class }$.
\end{itemize}

\begin{enumerate}
\item {\it The Hilbert-Schmidt estimate:}
from Lemma \ref{lemm:hspdo}, we get
\[ \| R_1 \|_{\rm H-S}=O( \ge^{-d/2}\ge ^{2\gg }|\log \ge | )~.\]
Using the fact that the Hilbert-Schmidt norm of
${\cal L}$ is of order $\ge^{-d/2}$, we  have
\[  \| R_2 \|_{\rm H-S}=O \left( \ge^{-d/2} \int _{T_\gg}^\infty 
e^{-2ks }ds \right) ~.\] 
Using  the explicit value $\gg = \gg_1 $, we get the result.

\item {\it The point-wise  estimate:}

\begin{lemm}
The Sobolev $H^s$ norm of a distribution $u$ in $X$ is, 
uniformly in $\ge$, equivalent to
 \[ \ge ^{-s}\| \hat{H}^{s/m}u    \|_{L^2}+\|u \|_{L^2}~.  \]
\end{lemm}
Taking $s>d$, using Sobolev's embedding Theorem 
on $X\times X$  and  the fact that
$\hat{H}$ commutes with $\Omega (s)$, we get the point-wise estimates
\[ [R_2](A,B)=O\left( \ge ^{-2s} \int_{T_\gg}^\infty \| \Omega (s+\tau)
\hat{H}^{s/m}{\cal L}\hat{H}^{s/m}\Omega ^\star (s) ds \|_{\rm H-S}ds \right) ~\]
and
\[ [R_2](A,B)=O \left( \ge^{-2s}\ge^{(1-2\gg)k/\Lambda } \right)~.\]
The result follows from the optimisation $\gg=\gg_2$.

\end{enumerate}

\end{demo}

\subsection{Applications: recovering the Hamiltonian from
the Correlations}

\begin{theo}\label{theo:recover_ham}
Let $U$ be the set of $(B,\xi_B)\in T^\star X $
so that there exists $t>0$ with 
$l(\Phi_{-t}(B,\xi_B) ,-H_0(B,\xi_B ))\ne 0$.
For any  $\tau _0 >0 $,
the restriction of the  Hamiltonian $H_0$ to $U$  can  be recovered 
from  the knowledge of $C(\tau )$ for $0<\tau <\tau _0$.
\end{theo}
\begin{rem} The physical meaning of the previous assumption
is that the ray dynamics and its Hamiltonian generator $H_0$ 
 can be recovered in the domain
where the rays cross, in the past time, the support of the
power spectrum of the source noise.
\end{rem}

\begin{demo} Let us assume, for simplicity,
 that $\pa ^2 H_0/\pa \xi ^2 $ is non degenerate. 
For $\tau >0  $ small enough, there is no
conjugate points on the unique trajectory from
$B$ to $A$ in time $\tau $ contained in $H^{-1}(I)$.
From Theorem \ref{theo:N=1} and Appendix B, we get
the following expression:
\[ C_{A,B}(\tau )=
(2\pi i \ge )^{-d/2}c(A,B,\tau )e^{iS(A,B,\tau ) /\ge }
+ R(A,B,\tau) \]
with
\begin{itemize}
\item $c(A,B,\tau )>0 $ if $(B,\xi_B)\in U $ (this follows from
  Section
\ref{ss:dyn_scalar} and the fact that the principal 
symbol $\pi $ of $\Pi $ is $>0$) 
\item The $L^2(X\times X)$  norm of $R(.,.,\tau) $
 is $O(\ge^{\ga -d/2})$ with $\ga >0$.
\end{itemize}
The Theorem follows: $S$ is a generating function 
for $\Phi_\tau $, the flow  $\Phi_\tau $ determines $H_0$
up to a constant. The constant can be extracted from
$S$ which is given as the action integral
$\int _{\gg}\xi dx -H_0 dt $ where $\gg $ is the trajectory
from $B$ to $A$ in time $\tau$. 

Without the simplifying assumption, the Theorem still holds: the H-S
norm of the ``Fourier integral operator''
$\Omega (\tau)\circ \opeps (\pi ) $ 
is of order $\ge^{-d/2}$.

\end{demo}

\section{The case of multi-component wave equations}
\label{sec:Ncomp}

\subsection{Wave equations: source white noise} \label{subsection:wave}
Let us take the case of  a scalar wave equation with constant
damping; the closest results were derived in \cite{WL3,RSK}.
We will make a formal calculation in the spirit of Remark
\ref{rem:green}.
We will consider  
 the  wave equation (\ref{equ:wave}), $u_{tt}+2a u_t -\Delta u= f$,
with 
\begin{itemize}
\item $a>0$ is a  constant damping coefficient
\item $\Delta $ a Riemannian laplacian in some Riemannian manifold $X$,
 possibly
with boundary:
\[ \Delta = g^{ij}(x)\pa _{ij}+ b_i(x)\pa _{i}\]
which is self-adjoint with respect to $|dx|$ and appropriate boundary
conditions; in fact we could replace the Laplacian by any 
self-adjoint operator on $X$!
\item $f=f(x,t)$ the source of the noise which will be assumed to be
a scalar white noise (homogeneous diffuse field):
\[ \E (  f(x,s)f(y,s') ) =\gd (s-s')\gd (x-y) \]
\end{itemize}

In order to get  a readable expression, it is convenient to introduce
$Q=\sqrt{-(\Delta + a^2)}$.
We get then easily:
\[ u(x,t)=\int _0 ^\infty ds \int_X  e^{-as}
\left[ \frac{\sin s Q}{Q}\right](x,y)f(y,t-s) |dy| \]
where $\sin sQ $ is defined from the spectral decomposition of $Q$
(any choice of the square root gives the same result for
$\frac{\sin s Q}{Q}$).
We define 
\[ G_a(t,x,y)=Y(t)\left[e^{-at} \frac{\sin t Q}{Q}\right](x,y)\]
with $Y$ the Heaviside function. We will call $G_a$ the {\it (causal)
 Green's function. } We can rewrite:
\[  u(x,t)=\int_{\R}ds \int_X  
G_a(t-s,x,y)f(y,s) |dy|~. \]

Let us assume now that $f$ is an homogeneous white noise and compute 
formally the correlation
${C_{A,B}(\tau )}$.

We get quite easily, using
\[ \sin \ga \sin \gb = \ha ( \cos (\ga - \gb) - \cos (\ga + \gb ))~:\] 

\begin{equation}\label{coralg}  {C_{A,B}(\tau )}=
\frac{e^{-a|\tau|}}{4a}\left[  (Q^2+a^2)^{-1}
\left( {\cos \tau Q}+
\frac{a \sin|\tau | Q}{Q}   \right)         \right](A,B) ~.
\end{equation}

Taking the $\tau $ derivative, we get the simpler formula:
\begin{equation}
 \frac{d}{d\tau }C_{A,B}(\tau )=\left\{
\begin{array}{l}\frac{-1}{4a}G_a(\tau, A,B)~{\rm for~}\tau>0\\
 \frac{1}{4a}G_a(-\tau, A,B)~{\rm for~}\tau<0 \end{array}
\right.
\end{equation} 
This ``exact'' relation 
between the correlation and the Green's
function in the case of a white noise source field
 is the starting point of many works in seismology. It has been
 derived
in various ways by several authors in particular R. Weaver
and O. Lobkis \cite{LW,WL1,WL3,RSK}. 

\begin{rem} The previous calculation can be extended
to the  more general wave equation (\ref{equ:wave_gen}):
\[ u_{tt}+ku_t -\Lambda u =f \]
with $k>0 $ constant, $\Lambda $ self-adjoint
and $f$ a white noise.
In particular, $C_{A,B}(\tau )$ is an {\bf even} function
of $\tau $.
This is the basis of
\begin{itemize}
\item Clock synchronisation \cite{SRTDHK}: if the clocks at the
  stations
$A$ and $B$ are not synchronised, the correlation is even
with respect to a shifted origin of times.
\item Test of applicability of the theory: in the semiclassical
regime studied later, approximate evenness holds. In fact the 
oscillating part is even because the corrections
to the Green's function do not affects the phases. In Theorem 
\ref{theo:N=1} the principal symbol $\pi $ is $\geq 0$.
This is basically due to the time reversibility of the ray dynamics.
\end{itemize}
\end{rem}

\subsection{General  wave equations: semiclassics}
\label{sec:generalwave}
\subsubsection{Assumptions}\label{sss:assumptions}
We want to derive results similar to those of Theorem \ref{theo:N=1}
in the case of   (multi-component) wave equations.
For that, we will introduce a matrix version of 
what is done in  Section \ref{sec:1comp} and use the results of
Section
\ref{ss:normal-multi}. Let us start with the following case of
Equation (\ref{equ:PI}):
\begin{equation} \label{equ:system}
 \frac{\ge}{i}{\bf u}_t +  \hat{H}{\bf u}=
 \frac{\ge}{i}{\bf  f} \end{equation}
where
\begin{itemize}
\item $ \hat{H}$ is a matrix of
 { $\ge-$pseudodifferential operators:}
\[   \hat{H} :=\opeps
(H_0+\ge H_1) \]
with (see Appendix A, Section \ref{ss:normal-multi}) 
\begin{itemize}
\item The {\bf principal symbol} $H_0(x,\xi):T^\star X\ra  {\rm
    Herm}(\C^N) $ belongs to $\Sigma _m^{\C^N,\C^N}$ and  is elliptic 
  of degree $m$ with  the eigenvalues
$\gl_1(x,\xi)<\cdots <  \gl_j (x,\xi)<\cdots  < \gl_n(x,\xi )$
 of constant multiplicities
$m_j$;
the corresponding eigenspaces $E_j(x,\xi) \subset \C^N$, of dimension
$m_j$,
are called the {\it polarisations.}
This assumption is satisfied for the elastic wave equation
in a manifold of dimension $3$  with
$N=6$: 2 eigenvalues $\pm \gl_S $ of multiplicity $2$ (the $S-$waves)
and 2   eigenvalues $\pm \gl_P $ of multiplicity $1$ (the $P-$waves).
\item The {\bf sub-principal symbol} $H_1(x,\xi)\in 
 \Sigma _m^{\C^N,\C^N}$  admits some
positivity property which controls the attenuation:
there exists $k>0$, such that the 
sub-principal symbols $H_{1,j},~j=1,\cdots, n,  $
satisfy
\[ \Im H_{1,j} \leq -k <0 ~.\]
\end{itemize}
\item The {\bf random field}
 ${\bf f}$ is given by ${\bf  f} =\opeps (l(x,\xi,\go)){\bf w}$:
\begin{itemize}
\item
 The noises ${\bf w}=(w_1,\cdots ,w_N)$ are independent
 white noises on $X\times \R$ 
\item 
 The matrix $l$ is smooth, compactly supported
w.r. to $(x,\xi) $ and its  Fourier transform w.r. to
$\go $ is compactly supported. The power spectrum of ${ f}$
is $(2\pi \ge)^{-(d+1)}ll^\star   (x,\xi,\go)$.
\end{itemize}
\end{itemize}

\subsubsection{The main result}

Using the gauge transform reducing $\hat{H}$ to the block diagonal
normal form of Section \ref{sss:normal},
 we can decompose the correlation into blocks
$C_{A,B}^{j,k}(\tau ),~j,k=1,\cdots, n$. 
\begin{theo}\label{theo:N>1}
With the  assumptions of Section \ref{sss:assumptions}
the correlations
$C_{A,B}^{j,k}(\tau )$ are given
\begin{itemize}
\item if $j=k$,  as in Theorem
\ref{theo:N=1} with $H_0=\gl_j$,
\item if $j\ne k$ and $l(x,\xi,\go) $ vanishes  for $x$  near $B$: 
$C_{A,B}^{j,k}(\tau )=O(\ge^\infty )$.
\end{itemize}
\end{theo}  
The proof makes a strong use of the normal form of Section \ref{sss:normal}:
we use a block decomposition of ${\cal L}={(\cal L}_{ij})$ in  Equation
(\ref{equ:cov_op}) and the block diagonal reduction
 $\Omega (t)=(\Omega _i(t) )$. We are
reduced to the study
of 
\[ \Omega _i(\tau)\int_0^\infty \Omega _i(t){\cal L}_{ij}\Omega
_j^\star (t)dt
~.\]
If $i=j$, we are in a situation very close to that of a scalar
Hamiltonian (Section \ref{ss:main-1}), while if $i\ne j$,
we can use Lemma  \ref{lemm:2dyn}.

In particular, for the elastic wave equation, if the source noise 
is supported away  from $B$, the correlation
$C_{A,B}^{P,S}$ 
between S-waves and P-waves is very small.


\section{Coda's correlations}\label{ss:coda}

In \cite{C-P}, M. Campillo and A. Paul computed the correlation
of {\it coda waves} and saw the emergence of the Green's function.
Let us try, rather informally, to discuss this observation, even
if we do not have mathematical results on it. 

Let $u(x,t)$ be the solution of the wave equation with  initially
localised Cauchy data (delta functions, no randomness
in the Cauchy data) at time $t=0$ (an "earthquake")
The coda is the field $u(x,t)$ in some window
$[T_1, T_2]$ with $T_1, T_2 >>0 $.
Let us look at the correlation
\[ C_{A,B}(\tau )=\lim _{T_1, T_2 \ra \infty}\frac{1}{T_2 -T_1}\int_{T_1}^{T_2}
 u(A,t+\tau )\bar{u}(B,t)dt ~\] assumed to exist.
Introducing the propagator $P$,
we get:
\[  C_{A,B}(\tau )=\int_X |dx| P(\tau, A,x)
\left(\lim _{T_1, T_2 \ra \infty}
 \frac{1}{T_2 -T_1}\int_{T_1}^{T_2} u(x,t )\bar{u}(B,t)dt \right) ~
.\]
So that the question is reduced to the study of the
 limits of $\frac{1}{T_2 -T_1}\int_{T_1}^{T_2} u(A,t )\bar{u}(B,t)dt $
as $T_1, T_2, T_2-T_1 \ra \infty$.
If we have some nice (smooth?) limit $K(A,B)$,
then $ C_{A,B}(\tau )=[\Omega (\tau)\circ  \hat{K}](A,B) $.

Looking at some semiclassical limit, this is closely related to
the following result by R. Schubert \cite{SCHU}: {\it if the classical
dynamics is hyperbolic and the time $t$ of the order
of the Ehrenfest time, the smoothed Wigner measures (``Husimi''
measures) of $u(t,.)$
are equidistributed on each energy shell.}
Does this uniformity of the energy distribution extend to
 the wave functions themselves, in particular to the phases?

\section{Using surface waves in order to image 
the inner crust} \label{sec:surface}

{\it \small In real applications to seismology, the fields are recorded
at the surface of the Earth. Moreover the main contribution
to the correlations comes from the surface waves:
\begin{itemize}
\item They have a smaller time decay than the body waves.
\item The source of the noise itself  (interaction with 
atmosphere and ocean) is located at the surface.
\end{itemize}
The Earth's crust acts as a {\rm  wave guide}.
This implies that we really have to look at a 2D version 
of the wave equation: the wave equation for surface wave is
an {\rm effective} wave equation on the surface. This equation
is  given in terms of eigenvalues of  a vertical 
Sturm-Liouville equation. Solving an {\rm inverse spectral problem}
for these Sturm-Liouville equation allows (in principle!) to image  the crust.
The inverse spectral problem is solved in \cite{YCdV5}.

In this Section, we will
discuss these effective Hamiltonians
in the  scalar model of the acoustic wave equation and show how one can 
reduce the problem of imaging the Earth's crust to an inverse spectral
problem. }

\subsection{A mathematical model}\label{sec:acoustic}

We  work locally in
 $X=\{ ({\bf x},z)\in \R^{2}\times \R~|~ z\leq 0 \} $
so that $z=0$ is the surface.
We will consider the very simple case of an acoustic wave equation
near the origin of $X$.
Let us give a function  
\[N({\bf x},Z):\R^{2} \times \R _- \ra \R _+ \]
which satisfies the following assumptions:
\begin{enumerate}
\item There exists $N_\infty >0$ and $Z_0 <0 $, 
so that $N({\bf x},Z)=N_\infty $ for $Z\leq Z_0$.
\item For every ${\bf x}\in \R^2$,
$\inf _{Z} N({\bf x},Z)=N_0 ({\bf x})>0 $
and $N_0({\bf x})< N_\infty $ for ${\bf x}$ close to $0$
\item $N$ is $C^1$ w.r. to $Z$ and the map
$ {\bf x}\ra N({\bf x},.)$,
 from 
$\R^2$ into the Banach space $ C^1 ([Z_0, 0],\R ) $,
is smooth.
\end{enumerate}

Let us introduce a small parameter $\ge >0$
and the following acoustic wave equation
where 
\begin{equation} \label{equ:n}
  n({\bf x},z)=N({\bf x},\frac{z}{\ge})~:\end{equation}

\begin{equation}\label{equ:ondes}
\left\{ \begin{array}{l}
u_{tt}-{\rm div}(n~{\rm grad}u)=0\\
\frac{\pa u}{\pa z}({\bf x},0)=0 
\end{array} \right.
\end{equation}

We plan to see that Equation~(\ref{equ:ondes}) admits, as $\ge \ra 0$,
asymptotic 
solutions of frequency of order
$\ge^{-1}$ located near the surface.
 Moreover these solutions are
determined by solving an effective pseudodifferential 
equation  on the surface  $\pa X= \R^2\times \{ 0 \}$.
In order to show that we can apply Theorem \ref{theo:normal-multi}
of Appendix A, we will 
first introduce the operator $H_0$ in
Section \ref{ss:sturm}. It will be a Sturm-Liouville
operator in the variable $Z$. 

\subsection{Discussion  of the assumptions
on the model}
The scalar acoustic wave equation (\ref{equ:ondes})  is a simplified
model of the elastic wave equation.
We could work with the elastic wave equation, but it will be more
complicated and the main ideas would be the same. 

 The function $n$ represent the square of the 
local propagation speed (group velocity). 
The formula (\ref{equ:n}) means that the speed admits fast variations
in the vertical directions and is smooth in the horizontal
direction. It is a plausible assumption that the the geological
structure near the surface of the earth is ``stratified''.

Assumption (1) on $N$ means that the speed is constant 
far from the surface. It is an  assumption
which could be removed at the price of a technical work.

Assumption (2) implies that the speed is somewhere smaller
than the speed at large depth. This kind of assumption is usually 
satisfied 
in seismology where the speed of elastic waves into  sediments 
layers is smaller than the speed inside the rocks below the sediments.

Assumption (3) means that we need less regularity in the 
vertical direction. It could be relaxed using piecewise
smoothness w.r. to $Z$, but at the price of a rather technical
work: the domain of the Sturm-Liouville operator
introduced below would then depend on ${\bf x}$.

\subsection{A family of Sturm-Liouville operators} 
\label{ss:sturm}

Let us consider, for each $({\bf x}, \xi) \in T^\star \pa X $,
 the self-adjoint differential operator
$L_{{\bf x} ,\xi }$ on the half line $Z\leq 0$, with
Neumann  boundary condition at $Z=0$, defined by:
\begin{equation} \label{equ:sturm}
L_{{{\bf x},\xi }}v:=-\frac{d}{dZ}
\left(N({\bf x},Z)\frac{dv}{dZ}\right) + N({\bf x},Z)| \xi | ^2 v ~.
\end{equation}
The domain $D\subset L^2(\R^-)$
of $L_{{\bf x} ,\xi }$ is 
\[ D:=H^2 (\R^-) \cap \{ v~|~v'(0)=0 \} \]
which is independent of $({\bf x} ,\xi )$. 

The spectrum of $L_{{\bf x} ,\xi }$ consists of a finite discrete spectrum
and a continuous spectrum $[ N_\infty | \xi | ^2,+\infty [$.
$L_{{{\bf x},\xi}}$ admits, for $\xi $ large enough,
 a non empty discrete spectrum
of simple eigenvalues
\[  N_0({\bf x})| \xi | ^2 < \gl _1 ({\bf x,\xi})<
\cdots < \gl _j  ({\bf x,\xi})<
\cdots  < \gl _k  ({\bf x,\xi})< N_\infty | \xi | ^2~,\]
which depend smoothly of $({\bf x},\xi)$.
In order to see that, we can interpret
  ${\cal L}_\hbar= | \xi |^{-2}  L_{{\bf x,\xi}}$
as a semiclassical Schr\"odinger type operator with an effective
Planck constant $\hbar =| \xi |^{-1}$ and a principal symbol
\[ p_{{\bf x }}(Z,\zeta)=
N({\bf x}, Z)(\zeta ^2 + 1) \]
 which admits a 
``well'' because of Assumption 2 in Section \ref{sec:acoustic}  on
$N$. 

We should however take care of  the fact that the number 
$k$ of eigenvalues  depends on $({\bf x,\xi})$ and goes to $\infty $
as ${\bf \xi}$ does. This will give birth to mode conversions
between surface waves and body waves. To my knowledge, such
mode conversions have not yet been studied from a mathematical
point of view.

\subsection{Wave guides as an adiabatic limit}
\label{ss:adia}

We will rewrite Equation (\ref{equ:ondes})
as an operator valued pseudodifferential equation 
as in Appendix A, Section \ref{ss:adia}, in the
spirit of the book \cite{Teufel}, Chapter 3 on
``space-adiabatic perturbation theory''.

Putting 
\[ u(t,{\bf x},z)=v(t,{\bf x},\frac{z}{\ge})\]
we get the following wave equation for $v$:
\[ \ge^2 v_{tt}-
\left( \pa _z \left( N \pa _z v \right)
+ \ge^2 N \Delta_{\bf x} v + \ge^2 {\rm grad}_{\bf x}N . {\rm grad}_{\bf x} v
\right)=0 ~,\]
which can be rewritten, using Weyl quantisation in the
variable ${\bf x}$:
\[ \ge^2 v_{tt}+ \opeps \left(
L_{\bf x,\xi} -i  \ge \xi .{\rm grad}_{\bf x}N \right) v= 0 ~. \] 
Putting $H_0({\bf (x,\xi)})=
L_{\bf x,\xi}$
and $H_1({\bf (x,\xi)}=-i \xi .{\rm grad}_{\bf x}N$, we 
will apply Theorem \ref{theo:normal-multi} in Appendix A.
Let us choose, for $({\bf x},\xi)$ close to $0$,
some eigenvalue 
$\gl_j (x,\xi) $ of $L_{\bf x,\xi}$
with a normalised eigenfunction
$\gf _j({\bf x}, \xi, Z)$. 
Then we have the existence of a propagation mode
of the form
\[ v(t,{\bf x},z) =w(t,{\bf x})\left( \gf _j({\bf x}, \xi, z/\ge)
+\ge \cdots  \right)\]
where $w$ satisfying the {\bf scalar} wave equation:
\[  \ge^2 w_{tt}+ \opeps (\gl_j (x,\xi)+\cdots ) w=0 ~.\] 

For each eigenvalue $\gl_j $, we can then 
apply the results of Section \ref{sec:Ncomp}.

From the passive imaging method, we can hope to recover the effective
Hamiltonians $\pm \sqrt{\gl_j} $.
Solving an inverse spectral problem for the Sturm-Liouville
operator $L_{\bf x,\xi}$ allows in principle to recover the function $n({\bf x},z)$
which encodes the physics of the propagation medium.

From the paper \cite{YCdV5}, we know that the function $N$ can be recovered
from the large $\xi$  asymptotic of the $\gl_j $'s.

\section{Appendix  A: a  review about
 pseudodifferential operators}\label{sec:pseudos}

{\it \small We review the basic definitions and properties of
  pseudodifferential operators (\OPD's).
Pseudodifferential operators (first called singular integral
operators
by Calder\'on and Zygmund) where developed in order to get
a class of operators generalising linear differential operators
and containing approximate inverses of elliptic operators. They became 
a basic tool in the theoretical study of linear differential 
operators with variable coefficients. A more general  theory was then
developed including a small parameter, here denoted by $\ge \in
]0,\ge_0]$.
Several textbooks are now available
\cite{YCdV1,Di-Sj,Du,Tr}. 
}

\subsection{Basic calculus}\label{sec:basic}
We will define the pseudodifferential operators ($\Psi$DO's) on $\R ^d$.
$\Psi$DO's on manifolds are defined locally by the same formulae.
\begin{defi}[Classical symbols]\label{defi:class-symb}
\begin{itemize}
\item 
The space $\Sigma _k $ of symbols of degree $k$
is the space of smooth functions
$p: T^\star \R^d \ra \C $ which satisfy
\[ \forall \ga ,\gb \in \N^d ,~ | D^\ga_xD^\gb _\xi p(x,\xi)|
\leq C_{\ga,\gb}\langle \xi \rangle ^{k}~,\]
with $\langle \xi \rangle =1+\|\xi \|$.
\item 
A {\rm classical} symbol of degree  $m$ 
  is a family of functions
\[ p_\ge : T^\star \R^d \ra \C \]
depending of some small parameter $\ge >0 $
which admits, as $\ge \ra 0$,  an asymptotic expansion
\[ p_\ge \equiv  \sum _{j=0}^\infty \ge ^{j} p_{j}(x,\xi) \]
with $p_j \in \Sigma _{m}$. We will denote this space by
$S^{m}_{\rm class}$.
\end{itemize}
\end{defi}

\begin{defi}[Pseudodifferential operators] \label{defi:opd}
A {\rm classical}  $\ge$-pseudodifferential operator $P_\ge$ (a $\Psi DO$)
of degree  $m$ 
on $\R^d$ is given  by the (distribution) kernel
\[ [P_\ge](x,x')=(2\pi \ge  )^{-d}\int _{\R^d}e^{i\langle  x-x'|\xi
 \rangle /\ge   }p_\ge \left(\frac{x+x'}{2}, \xi \right)
|d\xi |\]
where  $p_\ge (x, \xi) $, the so-called {\it (total) symbol} of $P$,
is in $S^{m}_{\rm class} $.

We will denote 
$ P_\ge=\opeps (p_\ge )$ and by $\Psi^m_{\rm class} $ the space of
classical  \OPD's of degree
$m$.
The operator $P_\ge$  is called the
{\rm Weyl-quantisation } of  $p_\ge$.
\end{defi}
The operators  $P_\ge $ act from  $C_o^\infty $
into $C^\infty $. 
The kernel of $P_\ge $ is  given by:
\begin{equation}\label{equ:quantif}
 [P_\ge ](x,x')=(2\pi \ge)
 ^{-d}\tilde{p_\ge}\left( \frac{x+x'}{2} ,\frac{x'-x}{\ge}\right)
 \end{equation}
with $\tilde{p}_\ge$ the partial Fourier transform of
 $p_\ge (x,\xi)$ w.r. to 
$\xi $.
Very often, one is only able to compute the symbol  $p_0$
which is called the {\it principal symbol} of ${P}$.
The term $p_1$ is called the {\it sub-principal} symbol.

The most basic fact about $\Psi DO$'s is the fact they 
form a (non-commutative) algebra:
if $P_\ge=\opeps(p_\ge)$ and $Q_\ge=\opeps (q_\ge)$, we have
$P_\ge Q_\ge=\opeps \left( p_\ge q_\ge +O(\ge) \right)$.
The composition formula for the total symbol is called  the {\it Moyal
$\star$-product}  $\opeps (a) \circ {\rm Op} (b)= \opeps (a\star b )$ and is given by  
\[ a\star b \equiv  ab +\frac{\ge}{2i}\{ a,b \}
+\sum _{j=2}^\infty \ge^j P_j(a,b) \]
where
\begin{itemize}
\item $ \{ a,b \}$ is the Poisson bracket
\item $ P_j(a,b)$ is a bi-linear bi-differential operator,
homogeneous of degree $j$ with respect to $a$ and $b$.
\end{itemize}

\subsection{The  symbols $S^{m}_\gd $}
In order to formulate the Egorov Theorem for long times, we will need
more sophisticated classes of symbols.

\begin{defi}[$S^{m}_\gd $]  For any real number
$m$ and any  $\gd$ with  $ 0\leq \gd< {\bf \ha} $, a symbol $a\in S^{m}_\gd $ 
is a smooth function on $T^\star X$ depending on $\ge$ which satisfies
\[ \forall \ga \in \N^{2d}, ~ \exists C_\ga >0,~\forall (x,\xi)\in T^\star X,
\quad |D^\ga_{x,\xi} a (x,\xi )|
 \leq C_\ga \ge^{-\gd |\ga |}\langle\xi\rangle^m ~.\]

\end{defi}
We have $S^{m}_{\rm class}\subset   S^{m}_0 $. 

We can associate to $a\in S^{m}_\gd  $ a pseudodifferential operator 
$\opeps (a)\in \Psi ^m_\gd $ using formula (\ref{equ:quantif}).
 Such operators  obey a nice  calculus,
see \cite{Di-Sj,Ev-Zw}:
if, for $j=1,2$, $a_j\in S^{m_j}_{\gd_j} $, $A_j :=\opeps
(a_j)$ and $\gd :=\max (\gd_1,\gd_2)$,
we have $A_1\circ A_2= \opeps (a_1 \star a_2)$
with $a_1\star a_2\in S^{m_1+m_2}_\gd $
and admits the asymptotic expansion given by  the Moyal
$\star -$product.
 We have $P_j(a_1,a_2)\in \ge ^{-j(\gd_1+\gd_2)} S^{m_1+m_2 }_\gd $.
In particular
$[A_1, A_2 ]=\frac{\ge}{i}\opeps \left( \{ a_1, a_2 \}
\right) + \opeps (b) $
with $b\in \ge^{2(1-\gd_1-\gd_2)} S^{m_1+m_2 }_\gd $. 

\begin{theo}[Calder\'on-Vaillancourt]\label{theo:cv}
The operators of  $\Psi^0_\gd $ extend to operators from 
$L^2 $ to $L^2$, uniformly continuous in $\ge$.
The operators in  $\Psi^m_\gd $ extends to operators from
$L^2 $ into   the Sobolev
space
$H^{-m}$ with a norm of order $O(\ge^m)$. 
\end{theo}

\subsection{Asymptotic expansions}

The basic tool in order to find a $\Psi DO$ is an inductive
construction based on the realisation of any asymptotic expansion:
given a sequence $a_j \in S^{m}_\gd $
and  $l_1=0 < l_2 < \cdots < l_j < \cdots $
and $l_j \ra +\infty $ as $j\ra \infty $, there exists
$a\in S^{m}_\gd $ such that 
\[ a\equiv  \sum _{j=1}^\infty \ge ^{l_j}a_j \]
meaning that
\[ \forall n,~  a-\sum _{j=1}^{n-1}\ge ^{l_j}a_j \in \ge^{l_n}S^{m}_\gd ~.\]
Such an $a$ is uniquely defined modulo a remainder term $r$
which lies in $\ge^\infty S^{m}_\gd $. The corresponding
\OPD~ $\opeps  (a)$ is well defined modulo 
$O(\ge ^\infty)$. This  implies, if $m\leq 0$  and  $a$  compactly
supported in $x$, that the $L^2\ra L^2$ norm of $\opeps (r)$
is $O(\ge^\infty)$.

\subsection{Microlocalisation}

A family $(u_\ge)_{\ge\to 0} $ of functions on $X$ is said to be
{\it admissible} if the $L^2$ norm on any compact set is at most of
polynomial
growth with respect to $\ge^{-1}$.

The family of functions $(u_\ge) $ is said to be $O(\ge^\infty )$ near a point
$(x,\xi)\in T^\star X$ if 
$\exists \gf \in C_o^\infty (X)$, $\gf (x_0)\ne 0$
and the $\ge-$Fourier transform
${\cal F}_\ge (\gf u_\ge)(\xi):=
(2\pi \ge )^{-d/2}\int e^{-i\langle x| \xi \rangle / \ge } u_\ge(x) dx
$
satisfies ${\cal F}_\ge (\gf u_\ge)(\xi)=O(\ge ^\infty )$
uniformly for $\xi$ close to $\xi_0$.

The micro-support ${\rm MS}(u_\ge)$  of $u_\ge $ is the set 
of all $(x,\xi)$'s at which $u_\ge$ is not $O(\ge ^\infty)$.
For any \OPD ~$A$, we have
${\rm MS}(Au_\ge)\subset {\rm MS}(u_\ge)$.

If $U$ is an open subset of $T^\star X$ and if $u_\ge$
and $v_\ge $ are admissible, we will shortly denote
$u_\ge =v_\ge +O(\ge^\infty) $ in $U$ if
${\rm MS}(u-v)\cap U=\emptyset $.

\subsection{Functional calculus and and an  useful  Lemma}
\label{ss:fc+ul}

A simple case of the functional calculus for \OPD's is as follows:
\begin{theo}\label{theo:fc}
Let $f\in C_o^\infty (\R)$ and
$A=\opeps (a_0 +\ge a_1 + \cdots )\in \Psi^m_{\rm class}$
with $a_0 $ elliptic (see Section \ref{sec:egorov}) and real. 
Then one can define $f(A)$ using the spectral theory
and $f(A)\in \Psi ^{-\infty }_{\rm class}$ with principal symbol $f(a_0)$.
\end{theo}
\begin{rem} The previous Theorem is usually stated  for $A$
self-adjoint  (see \cite{Sjo} Chap. 8). It can be  extended to the
case where the sub-principal symbol is not real, 
with a very close proof.
\end{rem}

\begin{lemm}\label{lemm:calL}
Let us consider the operator ${\cal L}$,
 given by
\[ {\cal L}:=\frac{1}{2\pi \ge}
\int \! \int  \Omega (-t)e^{it\go/\ge} \opeps^x  (L(x,\xi, \go))d\go dt~ \]
with $L\in S ^{-\infty }_{\rm class}(X\times \R)$ with $L(x,\xi, \go)$ having
 an $\go $-Fourier
transform
compactly supported as a function of $(x,\xi,t)$.
Here $\Omega (t) ={\rm exp}(-it \hat{H}/\ge )$ where 
$\hat{H}$ satisfies the assumptions of Section \ref{subsec:assumptions}. 
Then  ${\cal L}$  belongs to $\Psi^{-\infty}_{\rm class}$ and its  principal
symbol is given by 
$l_0(x,\xi)=L(x,\xi, -H_0(x,\xi))$.
\end{lemm} 
The integral giving  ${\cal L}$ is supported w.r. to $t$ in
the interval $[-t_0, t_0]$.
Using a change of variable $t=\ge t'$
in the Definition of ${\cal L}$, the proof is a simple corollary of
the functional calculus of \OPD's:
if $A=\opeps (a)$ with $a$ compactly supported,
the operator
$B={\rm exp}(it' \hat{H}) A $ belongs to $\Psi^{-\infty}_{\rm class}$
and the  principal symbol of $B$ is 
$b={\rm exp}(it'H_0) a $; because the function ${\rm exp}(it\cdot)$ is not
compactly
supported, we need first to rewrite
$A=F(\hat{H})A $ with $F\in C_o^\infty $
and then  use the functional calculus with
the function $f=F{\rm exp}(it\cdot)$.

\subsection{Hilbert-Schmidt estimate}
\begin{lemm}\label{lemm:hspdo}
If $P \in \Psi^{-\infty}_\gd (X)$ with $X$ a compact manifold, 
$P$ is Hilbert-Schmidt with 
$\| P \| _{\rm H-S}=O(\ge ^{-d/2}) $.
\end{lemm}
\begin{demo}
$P^\star P \in  \Psi^{-\infty}_\gd $ is trace
class (it has a smooth kernel) and its  trace  is easily
seen to be $O(\ge ^{-d}) $.
\end{demo}

\subsection{Multicomponent
operators}\label{ss:normal-multi}

\subsubsection{Systems of  \OPD }

\begin{defi}Let $E$ and $F$ be 2 Banach spaces
$\Sigma _m^{E,F}$
is the space of symbols of degree $m$ with values in the 
Banach space ${\cal L}(E,F)$ of continuous linear maps from $E$
into $F$. 
\end{defi}
Similarly, we can define the spaces
$S^{m, E,F }_{\rm class}$, $\Psi^{m, E,F }_{\rm class}$,
$\cdots $. The operators of  $\Psi^{m, E,F }_{\rm class}$
send functions on $X$ with values in $E$ into
functions on $X$ with values in $F$.

\begin{defi}
$H_0 \in \Sigma _m^{E,F}$ is {\rm elliptic}
if $H_0(x,\xi)$ is invertible for large $\xi$'s and the inverse
is in $\Sigma_{-m}^{F,E}$.
\end{defi}

Matrix valued \OPD's have properties similar to scalar \OPD's.

\subsubsection{Reduction}\label{sss:normal}

Let ${\cal H}$ be an Hilbert space and
$D\subset {\cal H}$ a dense sub-space.
\begin{defi}
A {\rm matrix valued Hamiltonian}
$H_0: T^\star X \ra {\rm Herm}({\cal H}) $
is an element of $\Sigma _m^{D,{\cal H} }$: for every $(x,\xi)\in T^\star X$,
$H_0(x,\xi)$ is a self-adjoint operator with domain $D\subset {\cal H}$.
\end{defi}
Let us consider a classical symbol 
$H_\ge \equiv H_0 +\ge H_1 + \cdots \in S^{m, D, {\cal H} }_{\rm class }$ 
with $H_0$ a matrix values Hamiltonian.
Let us consider near some point $z_0 \in  T^\star X$ an eigenvalue
$\gl_0 (z)$ of constant multiplicity $k$ and eigenspace 
$E(z)$ (the polarisation space)
and assume that $\gl_0 (z)$ is an isolated 
point in the spectrum of $H_0 (z)$.

Then we have, inspired from \cite{Em-We,Teufel}:
\begin{theo}\label{theo:normal-multi}
In some neighbourhood of $z_0 $, there exists
\begin{itemize} 
\item A $ {\cal L}({\cal H})-$valued \OPD ~
$\hat{U}={\rm Op}_\ge (U_0+\ge U_1 + \cdots )$
with $U_0(z) $ unitary.
\item A  classical symbol in $S_{\rm class}^{m,\C^k,\C^k}$
$\gl (z)=\gl_0 (z){\rm Id}+ \ge \gl_1 (z) +\cdots $
\item A symbol $K$ with values
in ${\cal L}({\cal D}\cap F,F )$
with $F=E^\perp $ with $\gl_0 (z) \notin {\rm Spectrum}(K_0(z) )$,
\end{itemize}
so that
\[ U^{-1} {\rm Op}_\ge({H}) U \equiv 
 \left( \begin{array}{cc} {\rm Op}_\ge({\gl })& 0 \\
0 & {\rm Op}_\ge({K}) \end{array} \right) \]
near $z_0$.

Moreover $\gl_1 (z)=\pi H_1 \pi + R_1 $
with $\pi $ the orthogonal projector onto
$E$ and $R_1$ self-adjoint.

\end{theo}
The proof is done by induction on the powers of $\ge$:
at the first step, we choose an unitary gauge transform
$U(z)\in{\cal L}({\cal H})$ so that
\[ U^{-1}H_0 U = \left( \begin{array}{cc}
\gl_0 {\rm Id} & 0 \\
0 &K_0 
\end{array}\right)  \]
with $\gl_0 {\rm Id} -K_0 $
invertible.
At the next step, we try to find
\[ B= \left( \begin{array}{cc}
0 & b \\
c & 0 
\end{array}\right)  \]
so that 
\[ (1-\ge B) \star \left( \left( \begin{array}{cc}
\gl_0 {\rm Id} & 0 \\
0 &K_0 
\end{array}\right)+ \ge \tilde{H}_1 \right)\star (1 +\ge B)=
\left( \begin{array}{cc}
\gl_0 {\rm Id} +\ge \gl_1 & 0 \\
0 & K_0 +\ge K_1  
\end{array}\right)+ O(\ge^2)~.\]
Collecting terms in $\ge ^1$, we get the following equations:
\[ b(\gl_0 {\rm Id} -K_0)=\gb ,~ (\gl_0 {\rm Id} -K_0 )c=-\gg \]
with 
\[ \tilde{H}_1=  \left( \begin{array}{cc}
\ga & \gb \\
\gc  &\gd 
\end{array}\right)     ~.\]
These equations can be solved with the assumption 
that $\gl_0 {\rm Id} -K_0$ is invertible.

\begin{rem} In the Theorem \ref{theo:normal-multi}, we do not
need that ${\rm Op}_\ge ({H})$ is self-adjoint (only $\opeps(H_0)$ is self-adjoint),
neither that  the symbol of ${\rm Op}_\ge ({H})$  takes values in bounded 
operators on ${\cal H}$.
\end{rem} 

\begin{rem} The Theorem \ref{theo:normal-multi}
will be used as well to reduce a multi-component wave equation
to scalar equation in Section \ref{sec:generalwave}
and as tool for adiabatic limits in
Section \ref{ss:adia}
\end{rem}

\begin{coro}\label{coro:blocs}
 Assuming that all eigenvalues
$\gl_1 < \cdots < \gl_j <\cdots <\gl_n$  of $H_0 (x,\xi) $ are of constant
  multiplicities
$m_j,~j=1,\cdots ,n$, the Hamiltonian $\hat{H}$
admits a micro-local normal form with diagonal blocks of
sizes $m_j \times m_j $
and principal symbols of the $j-$th block $\gl_j {\rm Id}_{\C^{m_j}}$.

The sub-principal symbol of the  $j-$th block is  
$H_{1,j}:= \pi_j H_1 \pi_j + R_j $ where $R_j $ is self-adjoint while
$\pi_j $ is the orthogonal projection on $E_j$; it means that, as
quadratic forms, $H_{1,j}$ is the restriction of $H_1$ to $E_j$. 
\end{coro}
The formula for the sub-principal symbols is a direct consequence
of \cite{Em-We}.

\subsubsection{Wave equations}

Let us consider the wave equation (\ref{equ:wave_gen})
\[ u_{tt}+Au_t -\Lambda u= 0 \]
with $\ge^2 \Lambda ={\rm Op}_\ge (H_0) $ ($H_0 $ elliptic)
and assume w.l.o.g. that $-\Lambda$ is $>0$ so that
$\Lambda _\ha = \ge \sqrt{- \Lambda}$ is also a \OPD.
Putting
\[ U:=\left( \begin{array}{c}\Lambda _\ha u \\ -i\ge u_t \end{array}
\right) \]
Equation (\ref{equ:wave_gen}) reads as:
\[ \frac{\ge}{i}U_t +\left(  \begin{array}{cc} O& -\Lambda _\ha \\
-\Lambda _\ha &-i\ge A  \end{array}
\right) U =0 ~,\]
which is of the form 
\[ \frac{\ge}{i}U_t+\hat{H} U=0 \]
with $\hat{H}$ an elliptic \OPD.

\subsubsection{Extension of Lemma \ref{lemm:calL} } 

Let us consider $a\in {\cal S}(\R_\go, L(\C^N))$
and $H_0 $ an $N\times N$ Hermitian matrix,
we define $a(H_0)$ as
\[ a(H_0):=\frac{1}{2\pi }\int e^{itH_0}{\cal F}a(t) dt ~.\]
The following Lemma is proved the same way as Lemma \ref{lemm:calL}.

\begin{lemm}\label{lemm:cal-L}
Let us consider the operator ${\cal L}$,
 given by
\[ {\cal L}:=\frac{1}{2\pi \ge}
\int \! \int  \Omega (-t)e^{it\go/\ge} {\rm Op}_\ge^x  (L(x,\xi, \go))d\go dt~ \]
with $L\in S ^{-\infty,\C^N,\C^N }_{\rm class}(X\times \R)$
 with $L(x,\xi, \go)$ having
 an $\go $ Fourier
transform
compactly supported as a function of $(x,\xi,t)$.
Here $\Omega (t) ={\rm exp}(-it \hat{H}/\ge )$ where 
$\hat{H}$ satisfies the assumptions of Section \ref{sec:generalwave}.
Then  ${\cal L}$  belongs to $\Psi^{-\infty,\C^N,\C^N}_0$ and his
principal
symbol
is given by 
$l_0(x,\xi)=L(x,\xi, -H_0(x,\xi))$.
\end{lemm}

\section{Appendix B: asymptotics of the Green's function
and classical dynamics}
\subsection{The scalar case}\label{ss:dyn_scalar}
We want to describe quite  explicitly the propagator $P(t,x,y)=
[\Omega (t)](x,y)$ with $\Omega (t)={\rm exp}(-it\hat{H}/\ge )$, 
$\hat{H}$ as in \ref{subsec:assumptions}.
We will do that in the energy interval $I$; it means
that we will only describe the action of $\Omega (t)$ on functions
whose micro-supports are  in $H^{-1}_0    (I)$.

To the Hamiltonian  $H_0:T^\star X \ra \R $, we associate the {\it ray dynamics}
defined by
\begin{equation}\label{equ:ham}
 \frac{dx_j}{dt}=\frac{\pa H_0}{\pa \xi_j},~
 \frac{d\xi_j}{dt}=-\frac{\pa H_0}{\pa x_j}~.\end{equation}

Let us denote by $\phi_t $ the flow of the Hamiltonian vector field
$X_{H_0} $ defined in Equation (\ref{equ:ham}).
The main result is that 
$P(t,.,.)$ is a {\it Fourier Integral Operator} associated to the flow
$\phi_t$. 
For simplicity, we will only describe the so-called {\it generic} case where
$P$ admits a WKB expansion.

If $x_0,y_0\in X$ and $\phi_t( y_0,\eta_0)=(x_0,\xi_0) $, we say that
$x_0$ and $y_0$ are {\it not conjugate}  along the ray
 $\gg(s)= \phi_s (y_0,\eta_0),~0\leq s\leq t$,
if the differential
$\gd \eta  \ra d\phi_t (0,\gd \eta ) $
 is invertible at the point $(y_0,\eta_0)$.
This implies that for $(x,y)$ close enough to $(x_0,y_0)$ there
exists an unique $(y,\eta )$ close to $(y_0,\eta_0)$
with $\phi_t (y,\eta)=(x,\xi)$.
In this case the action $S_\gg(t,x,y)$ is defined by
 $S_\gg(t,x,y)=\int _{\gg} \xi dx-H_0(x,\xi)dt $.
$S_\gg(t,.,.)$ is a {\it generating function} of $\phi_t$ near
$(y_0,\eta_0)$,  meaning that
$\phi_t (y,-\pa S_\gg/\pa y)= (x,\pa S_\gg/ \pa x) $
for $(y,\eta) $ close to $(y_0,\eta_0)$.

The action $S_\gg$ is useful in order to describe
a WKB expansion of the propagator $P(t,x,y)$ for $(x,y)$ close
to  $(x_0,y_0)$:
$P(t,x,y)$ is a sum of contributions $P_\gg (t,x,y)$ of rays going
from $y$ to $x$ in the time $t$.

Assuming that $x$ and $y$ are not conjugate along $\gg$, we have
\[ P_\gg (t,x,y) \equiv (2\pi  \ge)^{-d/2}\left( \sum _{j=0}^\infty 
\ge^j a_{j,\gg}(t,x,y) \right) e^{iS(t,x,y)/\ge} ~.\]
Moreover $a_{0,\gg}(t,x,y) >0$ if $t$ is small enough. 
\begin{defi} \label{defi:generic}
We will say that $(x,y,t)$ is {\rm generic} if 
 $x$ and $y$ are not conjugate along  any $\gg$ so that
$\gg(0)=x$, $\gg(t)=y $ and $\gg \subset H_0^{-1}(I)$.
\end{defi}
If  $(x,y,t)$ is generic, there is only a finite number of rays
from $y$ to $x$ in time $t$ included in $H_0^{-1}(I)$. 

 {\it Dispersive rays:} we say that the dynamics generated by 
$H_0$ is dispersive  
if $(H_0)_{\xi \xi }$ is invertible. This means that the map
$\xi \mapsto v(\xi):= \pa H_0 /\pa \xi $ ($v$ is the speed)
is a local diffeomorphism.
If this is the case, $(t,x,y)$ is generic for $x$ close enough to
$y$. 

{\it The dispersion relation:} the principal symbol of the wave
equation
is $\go + H_0(x,\xi):T^\star (X\times \R)\ra \R$.
 It is clear enough that the ray dynamics is
determined
by the hypersurface $\D :=\{ (x,\xi,\go)|\go + H_0(x,\xi)=0 \} $.
This surface is called the {\it dispersion relation.}
This terminology does not mean that $H_0$ is dispersive!

\subsection{The matrix case}\label{ss:dispersion}

Let us assume that $H_0$ is matrix valued as in Section \ref{sec:generalwave}.
The dispersion relation is then the hypersurface
\[ {\rm det}\big(\go {\rm Id }- H_0(x,\xi)\big)=
\D(x,\xi,\go)=0~, \]
  we can (at least outside a {\it mode conversion set}),
rewrite
\[ \D(x,\xi,\go)=\Pi _{j=0}^M \big(\go -\gl_j(x,\xi)\big)^{d_j}\]
which gives $M$ ray dynamics $\gl _{j}$ associated to different polarisations
\[ E_j=\ker \big(H_0(x,\xi)-\gl_{j}(x,\xi) {\rm Id }\big)~.\]

\section{Appendix C: long time Egorov Theorem}
The main reference  for this part is \cite{Bo-Ro} where the authors get
 only an operator norm estimates for the remainder. 
The proofs of  reference \cite{Bo-Ro} can be extended
in order
\begin{itemize}
\item to get  a point-wise result under some
ellipticity assumption
\item  to cover  the case of manifolds
\item to use a local Liapounov exponent:
the Liapounov exponent in some energy interval $I$
\item to allow an attenuation term.
\end{itemize}

\subsection{Egorov theorem}\label{sec:egorov}
\begin{defi}\label{def:admi}
 An {\rm admissible Hamiltonian} of degree $m>0 $ is $
\hat{H}:= {\rm Op}_\ge (H_0+\ge H_1)$
with
\begin{itemize}
\item $H_0\in \Sigma _m$  a real valued  function
 which is 
{\rm elliptic}:
\[ \exists c,\,C>0,~\forall (x,\xi)\in T^\star X ,\quad  H_0(x,\xi)\geq
C\langle\xi\rangle ^m -c ~.\]
\item  $H_1 \in \Sigma _{m-1}$ and $h_1=\Im H_1\leq -k  \leq 0 $
everywhere in $T^\star X$.
\end{itemize}
\end{defi}
\begin{exem}A typical admissible Hamiltonian
is the damped Schr\"odinger operator 
$\hat{H}=-\ge ^2 \Delta   +V(x)-i\ge a(x) $ with  $V(x)\geq 0 $
and $a(x)\geq 0$.
\end{exem}
We define the dynamics, for $t\geq 0$, by:
\[ \Omega (t)={\rm exp}(-it\hat{H}/\ge )~.\]
\begin{theo} [Egorov Theorem]\label{theo:egorov}
Let $\hat{H}=\opeps(H)$ be an admissible Hamiltonian and 
$A(0)=\opeps (a)$ with  $a \in C_o^\infty
(H_0^{-1}(I))$ with $I $ a compact interval. 
Then the operator $A(t)=\Omega (t)A(0)\Omega ^\star (t)$ 
is, for $t$ fixed,
 a pseudodifferential
operator whose symbol $a(t)\equiv \sum _{j=0}^\infty \ge^j a_j(t)$
 belongs to $ S^{-\infty}_{\rm class} $ 
and satisfies 
\begin{equation}\label{equ:ppal}
 a_0(t)(z)= {\rm exp}\left(2\int _{-t}^0 h_1 (\Phi_s (z)) ds\right)
a(\Phi _{-t}(z))~.\end{equation}
\end{theo}
Let us remark that the previous result is   a priory {\it uniform}
only in a fixed interval
$t\in [0,T_0]$ {\it independent of $\ge $.} 

Let us sketch the proof.
$A(t)$ is the value for $s=t$ of the solution $B(s)$ of the
ODE
\[  \frac{d}{ds}\Omega (t-s)B(s)\Omega ^\star (t-s) =0,~B(0)=A(0)~.\]

We have
\[ \frac{d}{ds}\Omega (t-s)B(s)\Omega ^\star (t-s) =
\Omega (t-s)\left( B'(s)+\frac{i}{\ge}(\hat{H}B(s)-B(s)\hat{H}^\star )\right)
\Omega ^\star (t-s)  ~.\]

Trying $B(s)={\rm Op}_\ge (b_0(s) +\ge b_1 (s)+ \cdots ) $
and 
using the Moyal product, we have to  solve
\[ b'(s)+\frac{i}{\ge}((H_0+\ge H_1)\star b(s) - b(s)\star (H_0+\ge
H_1))\equiv 0 ~.\]
We get the following equations by looking at the $\ge ^j$
term:
\[\frac{d}{dt}b_{j}+ \{ H_0 , b_j \}_{1}
-2 h_1 b_j =c_j ~,\]
with $c_0=0$ and, for $j\geq 1$,  $c_j=\sum _{k=0}^{j-1} P_{k,j} (b_k)$
where the $ P_{k,j}$'s are linear 
differential operators constructed by using the
Moyal product with $H_0$ and $H_1$.
If $\tilde{B}(s) ={\rm Op}_\ge (b_0(s) +\ge b_1 (s)+ \cdots )$,
we have
\[  \frac{d}{ds}\Omega (t-s)\tilde{B}(s)\Omega ^\star (t-s)=
\Omega (t-s) R(s) \Omega ^\star (t-s) \]
with $R(s)\in \ge^\infty \Psi ^{-\infty}_{\rm class}$.

It remains to prove that $\Omega (t-s) R(s) \Omega ^\star (t-s)$
is $O(\ge ^\infty )$ as an operator from the Sobolev space
$H^{-s} $ into the space
$H^s$, for any $s$.
From the Calder\'on-Vaillancourt Theorem
\ref{theo:cv}, it is enough to prove
that $L^2\ra L^2 $ norm of
$\hat{H}^{s/m}\Omega (t-s) R(s) \Omega ^\star
(t-s)(\hat{H}^\star)^{s/m}$ is $O(\ge^\infty)$.
This is clear because the powers of $\hat{H}$ (resp.
of $\hat{H}^\star$) commute with $\Omega (t-s)$
(resp. $\Omega ^\star (t-s)$ ).

\subsection{Long time Egorov Theorem}\label{sec:long}

We want to extend Egorov Theorem to times going
to infinity when $\ge\to 0 $.
From Lemma \ref{lemm:diff}, we get (see \cite{Bo-Ro}):
\begin{lemm}\label{lemm:estim-long}
Let $a_0 \in C_o^\infty (T^\star X)$ and $I= H_0  ({\rm Support}(a_0))$.
Let $T_\gg(I)$ be the time defined
in Definition \ref{defi:tgamma}. 
Then, the function 
$a_0(t)$ belongs to  $S^{-\infty }_{\gd }$ with $\gd =\ha-\gg  <\ha $ for all times $t\in [0, T_\gg(I  )] $,
and the constants in the
estimates of the derivatives are {\rm uniform in $t$}. 
Similarly, the symbolds of higher order
$\ge^j a_j $ belong to
$\ge^{j(1-2\gd)}S^{-\infty }_{\gd }$. 
\end{lemm}
The main result is:
\begin{theo}[Long time Egorov Theorem]\label{theo:long}
Let $\hat{H}=\opeps(H)$ be an admissible Hamiltonian (Section  \ref{sec:egorov})
and $a\in C_o^\infty (T^\star X)$.
For any $t\in [0, T_\gg] $ and with uniform symbol estimates
w.r. to $t$, 
 the operator $A(t):=\Omega(t)A\Omega^\star (t)$
 is a pseudodifferential operator,
the total symbol of which is given as in the classical Egorov Theorem
\ref{theo:egorov} by 
\[ A(t)\equiv \opeps \left(\sum_{j=0}^\infty \ge^j a_j(t) \right)\,, \]
where $\ge^j a_j(t)\in    \ge ^{j(1-2\gd)} S^{-\infty}_\gd $.

More precisely, for any $M$ and $k$, there exists $N$ so that
if $R_N(t)=A(t)-\opeps \left(\sum_{j=0}^N \ge^j a_j(t)\right)$,
the $C^k$ norm of the kernel of $R_N(t)$ satisfies, uniformly
for $t\in [0, T_\gg] $, the bound
\[ \| [R_N (t)] \|_{C^k(X\times X)}=O(\ge ^M)~.\]
\end{theo}

\begin{rem}
The  long time Egorov Theorem was first  proved by Bambusi, Graffi and Paul
\cite{Ba-Gr-Pa} and improved by Bouzouina 
and Robert \cite{Bo-Ro}.

 Our estimate of the remainder term is  better than the one given
in \cite{Bo-Ro} which is only an $L^2$ operator norm. We
need 
the ellipticity of $H_0$.
\end{rem}

Using Lemma \ref{lemm:estim-long}, the proof is the same as 
for a fixed time interval (see Section \ref{sec:egorov}).

\subsection{Integrals with two dynamics}\label{ss:2dyn}

\begin{lemm}\label{lemm:2dyn}
 Let us assume that we have two admissible Hamiltonians
$\hat{H}_\pm = \opeps (H_{0,\pm }+ \ge H_{1,\pm})$
with $k_\pm >0$.
Let us define 
the operator 
$$J:=\int_0^\infty \Omega _+(t+\tau )\opeps (a)  \Omega _-^\star (t)dt\,,$$
with $a\in C_o^\infty( T^\star X) $.
If we assume that, for $z\in {\rm Support}(a) $,
$H_{0,+}(z)\ne H_{0,-}(z)$ (no mode conversions occur in the support 
of $a$), then 
\[ J = \ge \Omega _+ (\tau ) K +R~,\]
with $K$ a \OPD ~of principal symbol
$k_0= -i a/(H_{0,+}- H_{0,-})$
and $R\in \ge^\infty \Psi ^{-\infty} $. 

In particular, if $a$ vanishes in some neighbourhood of $B$,
$[J](A,B)=0(\ge ^\infty )$. 
\end{lemm}

\begin{demo}
There exists some \OPD ~$K$ so that
$ \opeps (a)=-i(\hat{H}_+ K -K \hat{H}_-)^\star +r $
with $r \in \ge^\infty \Psi^{-\infty}$. 
$J$ can be rewritten as
\[J= \ge \int _0^\infty \frac{d}{dt}
\left( \Omega _+(t+\tau )K   \Omega _-^\star (t)\right) dt 
+\int  _0^\infty
\left( \Omega _+(t+\tau ) r  \Omega _-^\star (t)\right) dt   ~.\]
Integration by part gives the result.
\end{demo}

\section{Appendix D: stationary Gaussian fields are ergodic}
\label{app:gaussian}
This section is inspired from \cite{BGP}.
Let us define
\[ C_T (\tau , A,B):=\frac{1}{T}\int_0^T u(A,t)\otimes \overline{
u(B,t-\tau) }dt ~,\]
where $u$ is the causal solution of Equation(\ref{equ:PI}). 
We then have 
\begin{theo}
Let us assume that $f(.,t)=Kw(.,t)$ where $w(x,t)$
is a (Gaussian) white noise on $X\times \R $
and $K:{\cal H}\ra {\cal H}$.
Then 
\begin{enumerate}
\item If $K$ is Hilbert-Schmidt, $C_T(\tau , A, B) \ra
 \E ({C(\tau
    ,A,B)}) $,
as $T\ra \infty $, almost surely for almost every pair
$(A,B)$.
\item If $K$ is smoothing and $\hat{H}$ elliptic,
 $C_T(\tau , A, B) \ra \E ({C(\tau
    ,A,B)}) $, as $T\ra \infty $, almost surely
 in the $C^\infty (X\times X)$
topology.
\end{enumerate}
\end{theo}
\begin{demo}
1. We will assume w.l.o.g. that $u$ is scalar valued and real and
 use the following classical identity:
if $f=(f_1, f_2, f_3, f_4):\Omega \ra \R^4$
is a Gaussian random variable,
\begin{equation}\label{id:gauss}
 \E ( f_1f_2f_3f_4) =
 \E ( f_1 f_2 )  \E ( f_3 f_4 ) +
 \E ( f_1 f_3  ) \E ( f_2 f_4  ) +
 \E ( f_1 f_4  )  \E (  f_2 f_3 ) ~.\end{equation}

We have 
\[ \E ( \| C_T (\tau )\|_{\rm H-S}^2 )=
\frac{1}{T^2}\int_0^T dt \int _0^T dt'\int _{X\times X}dA dB 
 \E ( u(A,t)u(B,t-\tau ) u(A,t')u(B,t'-\tau )) ~.\]
We apply the identity (\ref{id:gauss})
and get:
\[ \E ( \| C_T (\tau )\|_{\rm H-S}^2 )=
\|  \E (  C_T (\tau ) ) \|_{\rm H-S}^2 +
\frac{1}{T^2}\int_0^T dt \int _0^T dt'\int _{X\times X}dA dB (II+III)
~,\]
with 
\[  II=  \E ( u(A,t)u(A,t' ))
 \E (  u(B,t-\tau)u(B,t'-\tau )) \]
and
\[ III= \E ( u(A,t)u(B,t'-\tau  ))
 \E (  u(A,t-\tau)u(B,t' )) ~. \]

For example, we have
\[  \E (  u(A,t)u(B,t'-\tau ))=
[ \int _0^\infty ds \int_0^\infty ds'
\Omega (s) K K^\star \Omega ^\star (s') \gd (t-s=t'-s'-\tau )](A,B) \] 
whose $L^2$ norm as a function of $(A,B)$
 is estimated as $O({\rm exp}(-|t-t'+\tau |))$.
Using Cauchy-Schwarz inequality and the fact
that a product of two Hilbert-Schmidt operators
is trace class, we get
\[ |\frac{1}{T^2}\int_0^T dt \int _0^T dt'\int _{X\times X}dA dB (II+ III)|
=O(1/T)~. \]
Standards argument involving Tchebichev inequality allow to conclude.

2.
The second assertion is proved in a similar way using the Sobolev
embeddings, like
at the end of the proof of Theorem \ref{theo:egorov}.
\end{demo}

\bibliographystyle{plain}

\end{document}